%% file: main.tex
\documentclass[a4paper,fleqn]{cas-dc}

\usepackage{amssymb}
\usepackage{array}
\usepackage{tabularx}
\usepackage{float}
\usepackage{amsmath}
\usepackage{microtype}
\usepackage{graphicx}
\usepackage[numbers]{natbib}
\usepackage[utf8]{inputenc}
\usepackage[T1]{fontenc}
\usepackage{stfloats}
\usepackage{url}
\usepackage{siunitx}
\usepackage{url}
\usepackage{hyperref}
\usepackage{enumitem}
\usepackage[version=4]{mhchem}
\usepackage{caption}
\usepackage{makecell}
\DeclareSIUnit\angstrom{\text{Å}}

\def\tsc#1{\csdef{#1}{\textsc{\lowercase{#1}}\xspace}}
\tsc{WGM}
\tsc{QE}
\tsc{EP}
\tsc{PMS}
\tsc{BEC}
\tsc{DE}

\begin{document}
\let\WriteBookmarks\relax
\def\floatpagepagefraction{1}
\def\textpagefraction{.001}
\shorttitle{Thin-Fim Solar Photovoltaic: Trends \& Futures}
\shortauthors{Donald Intal \& Abasifreke U. Ebong}

\title [mode = title]{Thin-Film Solar Photovoltaics: Trends and Future Directions}                      
\author[1]{Donald Intal}[type=editor,
                        auid=000,
                        bioid=1,
                        orcid=0000-0003-3528-4894]
\fnmark[*]
\ead{dintal@charlotte.edu}

\affiliation[1]{organization={Department of Electrical and Computer Engineering, University of North Carolina at Charlotte},
                addressline={9201 University City Blvd}, 
                city={Charlotte},
                postcode={28223}, 
                state={North Carolina},
                country={USA}}

\author[1]{Abasifreke U. Ebong}[type=editor,
                        auid=000,bioid=1,
                        orcid=0000-0002-8808-5461]

\ead{aebong1@charlotte.edu}

\credit{Data curation, Writing - Original draft preparation}

\cortext[cor1]{Corresponding author}

\begin{abstract}
Thin-film photovoltaic (PV) technologies address crucial challenges in solar energy applications, including scalability, cost-effectiveness, and environmental sustainability. This paper reviews critically, thin-film technologies such as amorphous silicon (a-Si), cadmium telluride (CdTe), and copper indium gallium selenide (CIGS). It also discusses emerging technologies, including perovskites, copper zinc tin sulfide (CZTS), quantum dots (QDs), organic photovoltaics (OPV), and dye-sensitized solar cells (DSSC). Among these, CdTe and CIGS currently dominate commercial viability, achieving laboratory-scale efficiencies of 23.1\% and 23.6\%, respectively. Perovskites have notably advanced, reaching a laboratory efficiency of 26.7\%. Thin-film PV technologies significantly reduce material use and manufacturing costs, offering distinct advantages such as flexibility and lightweight structures, thereby enabling diverse applications from building-integrated systems to portable electronic devices. Despite these benefits, broader adoption remains limited by challenges including long-term stability, toxicity concerns, and material scarcity. Addressing these challenges through advancements in tandem architectures, improved encapsulation strategies, and sustainable material sourcing is essential for thin-film PV technologies to substantially contribute to the global renewable energy transition.
\end{abstract}


\maketitle

\input{Introduction}

\input{Traditional}
\input{Emerging}

\input{CommercialModule}

\input{Market}

\input{Conclusion}

\input{References}





\end{document}

%% file: Introduction.tex
\section{Introduction}
Tackling the twin imperatives of mitigating climate change and safeguarding energy security demands a rapid, large-scale deployment of renewable technologies.  Among these, solar photovoltaics (PV) stand out for their near-unlimited resource base, falling levelized cost of electricity (LCOE), and modular scalability from milliwatt sensors to multi-gigawatt utility parks.  Crystalline silicon (c-Si) has long dominated the PV market, setting benchmarks for efficiency and reliability.  Yet the technology’s strengths come at the expense of energy- and material-intensive production: Czochralski crystal growth operates above 1500 $^\circ$C; diamond-wire sawing wastes 35$-$40 \% of the ingot as kerf; and cell processing still relies on high-vacuum metallization and phosphorus diffusion furnaces \cite{hermle2020passivating}.  Moreover, single-junction c-Si is fundamentally capped by the 33.7 \% Shockley$-$Queisser limit \cite{ehrler2020photovoltaics}, making further efficiency gains increasingly incremental and expensive.

Multi-junction architectures provide an avenue past this limit.  III$-$V triple-junction cells already exceed 39 \% under one-sun conditions, but their epitaxial growth and scarcity of Ga and In constrain mass deployment.  By contrast, perovskite/silicon tandems combine a wide-gap perovskite (1.7$-$1.8 eV) with a 1.1 eV c-Si bottom cell, retaining much of the existing silicon manufacturing infrastructure while pushing certified efficiencies to 27 \% and roadmap targets beyond 30 \% by 2030 \cite{yamaguchi2021multi,kamaraki2021perovskite,wu202227}.

\begin{figure}[!ht]
    \centering
    \includegraphics[width=\columnwidth]{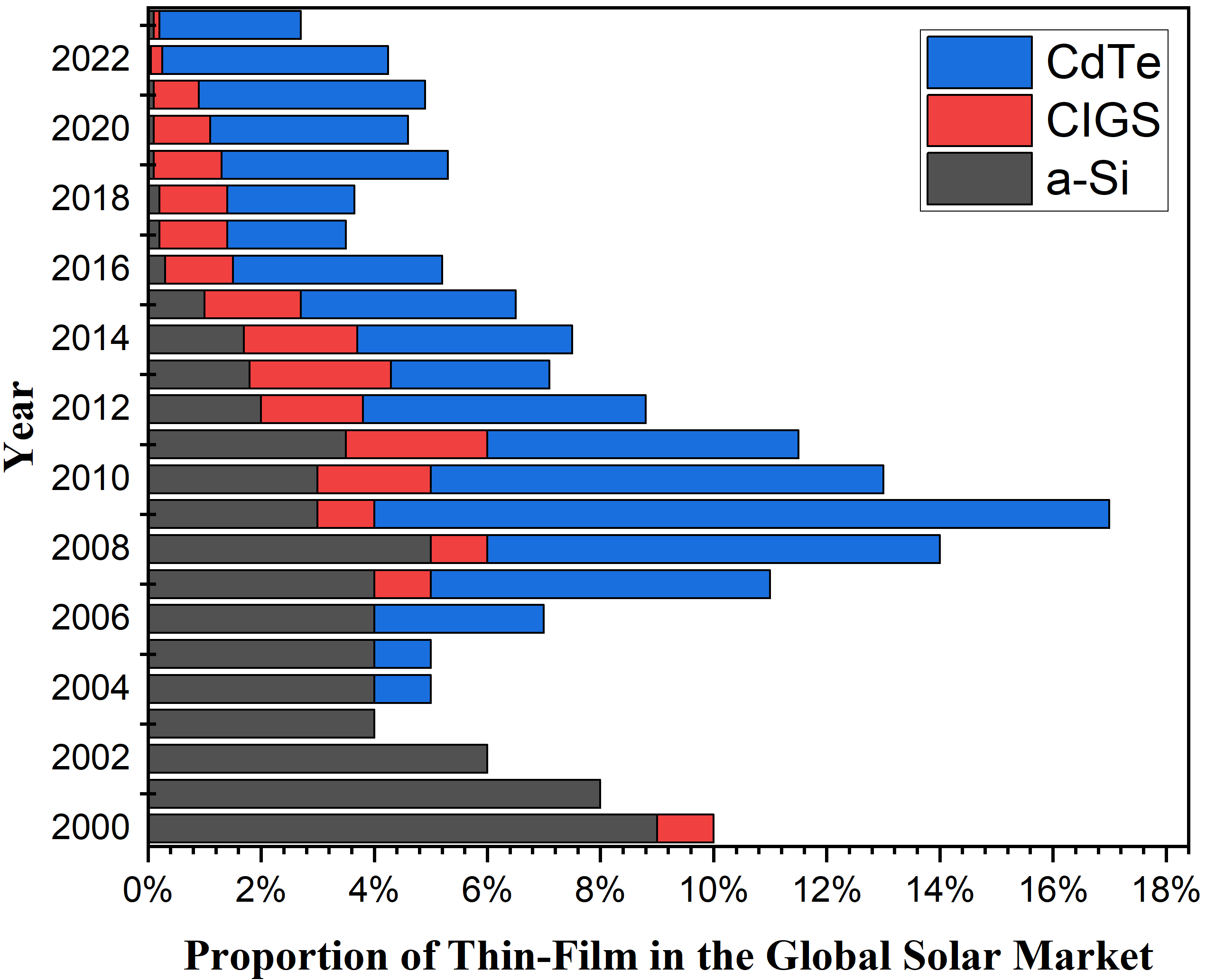}
    \caption{Market share of thin-film technologies from 2000 to 2023 \cite{FraunhoferISE2024,ITRPV2024}.}
    \label{fig:marketshare}
\end{figure}

Thin-film solar cells offer a complementary route that replaces 160 $\mu$m wafers with 1$-$3 $\mu$m absorbers deposited on glass, metal foil, or polymer.  This geometry slashes semiconductor usage by >95 \%, enables continuous roll-to-roll (R2R) or sheet-to-sheet processing, and unlocks form factors unreachable with brittle wafers.  Applications now span building-integrated photovoltaics (BIPV), vehicle and drone surfaces, agrivoltaic shade nets, and space-qualified ultralight modules \cite{FraunhoferISE2024,ramanujam2020flexible}.  Advances in slot-die coating, laser-scribing interconnects, and atmospheric-pressure deposition have pushed factory throughput above 15 m min$^{-1}$ while holding capital expenditure below \$0.15 W$^{-1}$ \cite{hambach2022laser}.  As depicted in Figure \ref{fig:marketshare}, the combined market share of CdTe, CIGS, and \textit{a}-Si peaked at ~15 \% in 2010, dipped to ~5.5 \% in 2024, and is forecast to rebound as next-generation thin films mature.

Each thin-film technology faces distinct hurdles.  CdTe cells rival mid-range c-Si in efficiency yet rely on a toxic element subject to stringent disposal rules.  CIGS offers high performance and flexibility, but indium and gallium criticality questions cloud scalability.  Perovskites have raced past 26.7 \% in the lab, yet outdoor stability and lead management remain unresolved. \textit{a}-Si:H suffers from Staebler–Wronski light-induced degradation, capping practical efficiency around 9 \%.  Organic, quantum-dot, and dye-sensitised devices promise unique niches transparency, colour tuning, indoor illumination but must overcome longevity and encapsulation challenges.

\begin{figure}[!ht]
    \centering
    \includegraphics[width=\columnwidth]{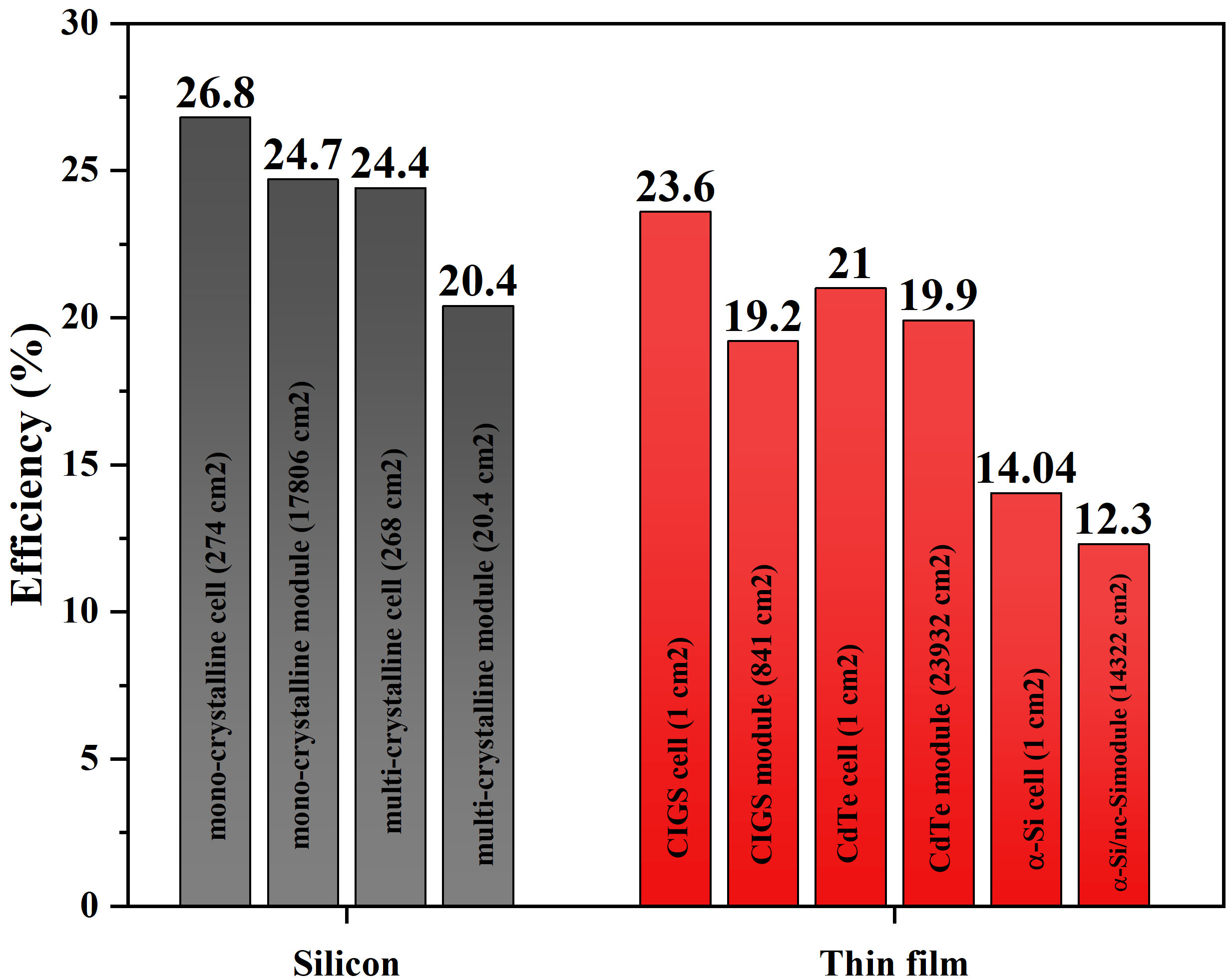}
    \caption{Efficiency comparison of the best laboratory cells with a minimum area of 1 cm$^{2}$ and the best commercial modules \cite{green2024,nrel2023,sai2016stabilized}.}
    \label{fig:eff1}
\end{figure}

The 2024 International Technology Roadmap for Photovoltaics (ITRPV) projects continued incremental gains for mainstream c-Si (TOPCon, heterojunction, IBC) alongside breakthrough potential in thin-film domains (Figure \ref{fig:eff1}).  Notably, tandem stacks that marry thin-film wide-gap absorbers with silicon or CIGS sub-cells are slated to push module efficiencies beyond 30 \% while shaving grams-per-watt material intensity.

This paper examines the potential of thin-film solar cells as scalable and cost-effective alternatives to crystalline silicon technologies. A detailed comparison of their performance, costs, and market potentials is provided. Additionally, the paper explores current innovations, key challenges, and future research directions, emphasizing the role of thin-film solar cells in shaping the global renewable energy landscape.

%% file: Traditional.tex
\section{The Traditional Thin Film Technologies}

\subsection{Amorphous Silicon}
Amorphous silicon (\textit{a}$-$Si) has a direct optical band-gap of
\(E_g \approx 1.7\ \text{eV}\), so its absorption coefficient exceeds
\(10^{5}\ \text{cm}^{-1}\) across most of the visible spectrum.
Consequently, more than 90\,\% of incident sunlight can be absorbed in a
film only 300$-$500 nm thick, enabling ultrathin devices and roll$-$to$-$roll
manufacturing on low-cost substrates such as stainless steel or polymer
foils \cite{chopra2004thin}.
The random Si$-$Si network, however, contains
\(\sim 10^{20}\ \text{cm}^{-3}\) dangling-bond defects (mid$-$gap
\textit{D} states) that pin the Fermi level, restrict minority$-$carrier
diffusion lengths to well below 0.5 $\mu$m, and depress
the mobility–lifetime product to the
\(10^{-9}\text{–}10^{-8}\ \text{cm}^2\text{V}^{-1}\) range several orders
of magnitude poorer than in crystalline silicon.

During plasma$-$enhanced chemical$-$vapor deposition (PECVD) at
200$-$300 \textdegree C, atomic hydrogen passivates most dangling bonds, forming predominantly monohydride (Si$-$H) and dihydride (Si$-$H\(_2\)) configurations.  This reduces the defect density to
\(10^{15}\text{–}10^{16}\ \text{cm}^{-3}\), improves photoconductivity by two to three decades, and extends carrier diffusion lengths to 1$-$2 $\mu$m in high-quality intrinsic (\textit{i}) layers \cite{chopra2004thin}.  The improvement is partly offset by the Staebler$-$Wronski effect (SWE): light-induced breaking of Si$-$H bonds creates new dangling bonds, leading to a 15$-$20 \% efficiency loss within the first 1000 h of operation.  A short anneal at 150$-$200 \textdegree C heals the damage, but long-term mitigation still requires carefully optimised deposition conditions.  Strategies include (i) lowering ion$-$bombardment energy, (ii) inserting nanocrystalline “seed” layers to relieve internal strain, and (iii) tuning the H dilution ratio to favour stable monohydride bonding.

\begin{figure}[!ht]
   \centering
    \includegraphics[width=.8\columnwidth]{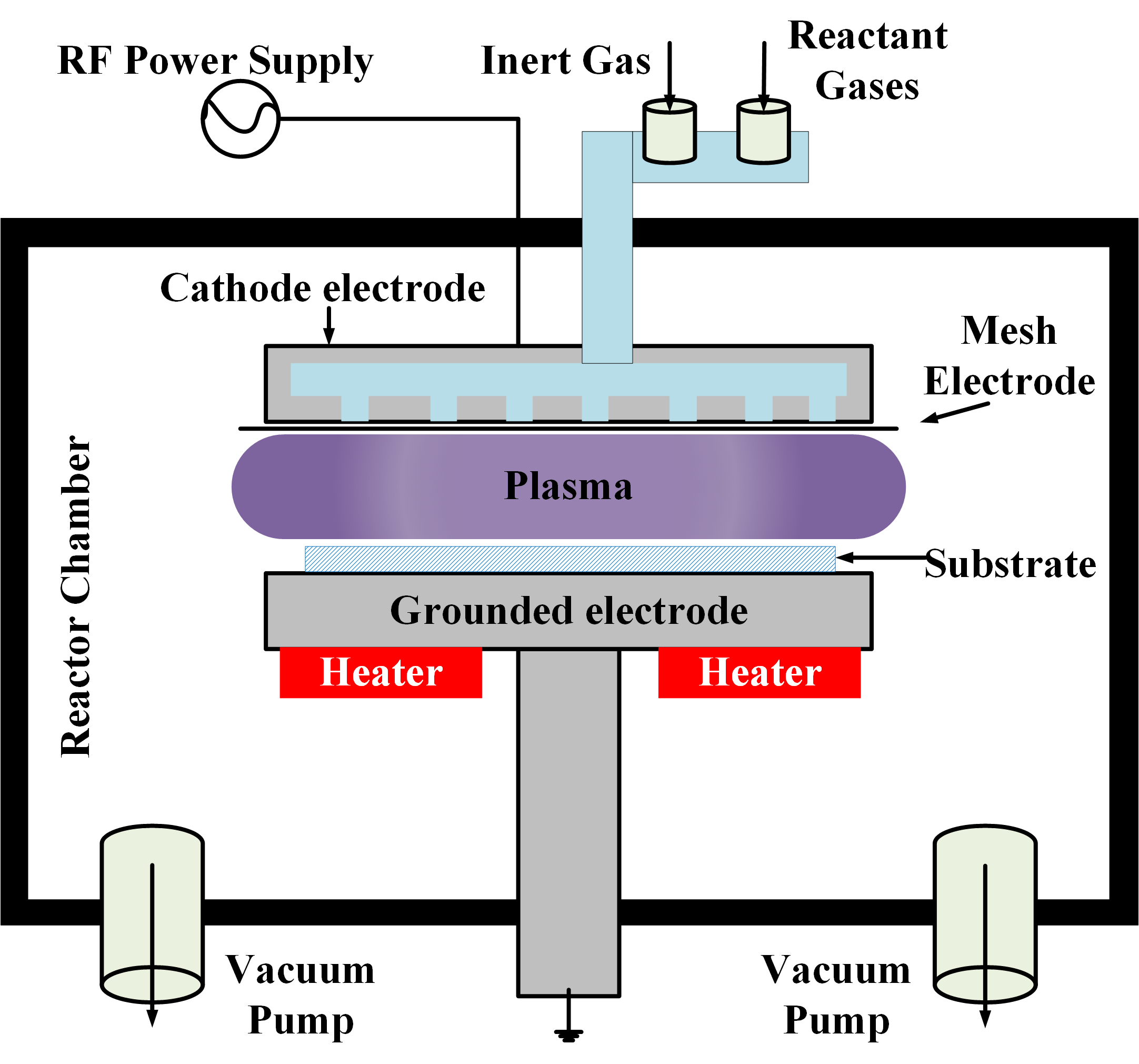}
   \caption{Diagram of Triode PECVD setup: A mesh electrode is introduced between the cathode and teh plasma region, enhancing control over plasma distribution and improving thin$-$film deposition quality by reducing ion bombardment on the substrate.}
    \label{fig:triode}
\end{figure}

The \emph{triode} PECVD reactor shown in Figure~\ref{fig:triode} decouples plasma generation from film growth: a remote RF electrode sustains the discharge, while a third grid extracts a low-energy ion flux towards the substrate.  This geometry suppresses plasma damage and allows thicker, low-defect \textit{i}$-$layers (\(>500\ \text{nm}\)) that absorb more red and near$-$infrared photons without exacerbating SWE.

Commercial single$-$junction \textit{p$-$i$-$n} \textit{a}-Si:H modules stabilize at 9$-$10 \% efficiency.  Higher performance is obtained with \emph{band-gap-graded multijunction} stacks, whose sub-cell parameters are summarized in Table~\ref{tab:aSi_multijunc}.  Careful current matching among the three junctions yielded the long-standing 13.6 \% record \cite{kim2013remarkable,sai2015triple}; applying triode PECVD to the middle and bottom sub-cells, combined with improved doped-layer conductivity and enhanced back reflectors, pushed the figure to \textbf{14.0 \%}—still the highest independently confirmed efficiency for an all-\textit{a}-Si:H triple-junction device as of 2025 \cite{sai2016stabilized,Lee2017A}.

\begin{table}[h!]                           
  \footnotesize                            
  \setlength{\tabcolsep}{4pt}              
  \caption{Typical absorber materials and stabilised short-circuit
           current densities in a high-efficiency \textit{a}-Si:H
           triple-junction cell.}
  \label{tab:aSi_multijunc}
  \centering
  \begin{tabular*}{\columnwidth}{@{\extracolsep{\fill}}lccc}
    \hline
    \textbf{Junction} &
    \textbf{Absorber} &
    $E_g$ (eV) &
    $J_{\mathrm{sc}}$ \\[-4pt]
    &  &  & (mA\,cm$^{-2}$)\\
    \hline
    Top    & \textit{a}-Si:H                    & 1.85–1.90 & 4–5  \\
    Middle & \textit{a}-SiGe:H (~20 \% Ge)      & 1.55–1.65 & 6–7  \\
    Bottom & $\mu$c-Si:H or \textit{a}-SiGe:H   & 1.10–1.30 & 10–12\\
    \hline
  \end{tabular*}
\end{table}

Thanks to its low deposition temperature, the energy-payback time of
\textit{a}-Si:H PV modules remains below 1.5 years even in temperate
climates, and their specific energy yield (kWh kWp\(^{-1}\)) is
competitive under diffuse-light conditions.  Current research targets
SWE-resilient hydrogen bonding configurations, advanced light-management schemes, and hybrid \textit{a}-Si:H/organic or \textit{a}-Si:H/perovskite tandems, with the goal of approaching a stabilized efficiency of 18 \% without sacrificing the low-temperature, low-cost manufacturing advantages that first made amorphous silicon attractive.

\subsection{CIGS}

\begin{figure}[!ht]
    \centering
    \includegraphics[width=\columnwidth]{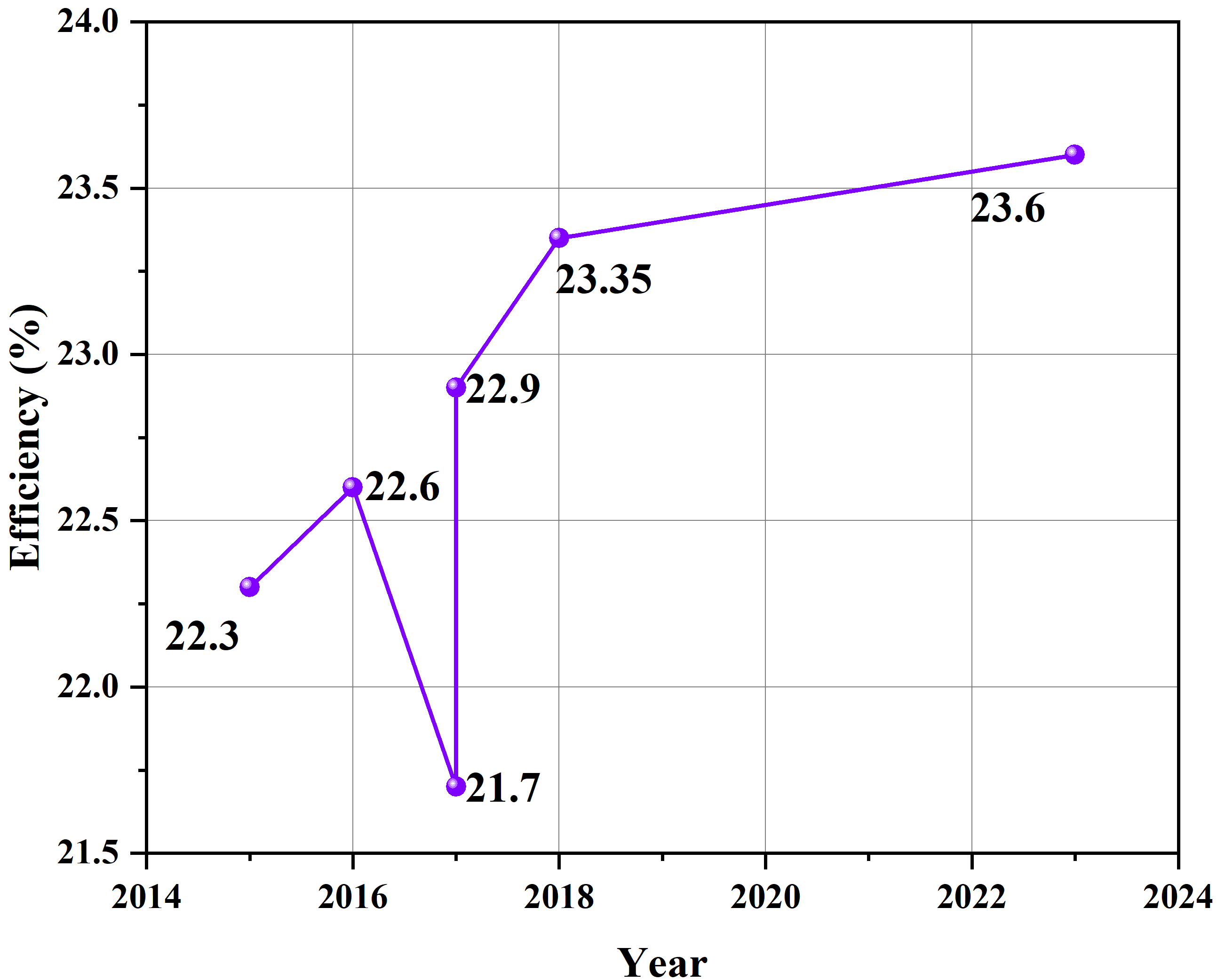}
    \caption{Best laboratory efficiencies for CIGS from 2015 to 2024.}
    \label{fig:cigseff}
\end{figure}

The first CuInSe$_2$ (CIS) thin-film solar cell, reported by Kazmerski in 1976, used a vacuum-evaporated, \emph{Cu-rich} absorber ($\mathrm{Cu}/(\mathrm{In}+\mathrm{Ga})>1$) on glass held at $\sim 400\,^{\circ}\mathrm{C}$.
Abundant Cu$_2$Se secondary phases and uncontrolled grain-boundary defects limited the open-circuit voltage, \(V_\mathrm{OC}\), to only 0.18 V, so the device reached a mere \textbf{4.5 \%} efficiency \cite{kazmerski1976thin}. Follow-on composition studies showed that mildly \emph{Cu-poor} films ($0.80\!\lesssim\!\mathrm{Cu}/(\mathrm{In}+\mathrm{Ga})\!\lesssim\!0.95$) suppress those secondary phases and promote large-grained chalcopyrite growth. Adding a 50$-$60 nm CdS buffer layer and a ZnO/Al:ZnO (AZO) transparent-conducting bilayer formed a type-II heterojunction that repelled electrons from the interface; together with a grain-enlarging, three-stage co-evaporation process, these refinements raised the record to \textbf{11.2 \%} in 1985.

Substituting part of In with Ga to create
Cu(In$_{1-x}$Ga$_x$)Se$_2$ (CIGS) widens the optical band gap from 1.04 eV to 1.2$-$1.3 eV, enabling higher \(V_\mathrm{OC}\) and reducing thermalization losses. The champion \textbf{19.9 \%} cell reported in 2008 employed a \emph{band-gap-graded} absorber: a Ga-poor “notch’’ (0.9$-$1.0 eV) near the CdS interface to avoid a conduction-band cliff, grading to a Ga-rich layer (1.4$-$1.5 eV) at the Mo back contact to reflect minority carriers and cut rear-surface recombination \cite{repins200819}. Sodium diffusing from the soda-lime glass substrate further passivated In$_\mathrm{Cu}$ antisite defects and boosted the hole density, driving the \(V_\mathrm{OC}\) deficit, \({E_g}/{q}-V_\mathrm{OC}\), below 0.40 V, the best result of its era.

In the 2010s two breakthroughs, \textit{alkali post-deposition treatments} (PDTs) and \textit{flexible polymer substrates} pushed
single-junction efficiencies beyond 21 \%. Ex-situ KF-PDT produces a thin KIn$_{1-y}$Ga$_y$Se$_2$ surface layer (\(E_g > 2\) eV) that electrostatically repels electrons from the heterointerface, while K$^+$ ions at grain boundaries neutralize dangling bonds. Using polyimide foils equipped with an engineered Na reservoir, Jackson~\emph{et al.} demonstrated a lightweight, rollable device with \textbf{21.7 \%} efficiency \cite{jackson2015properties}. Lee and Ebong’s review lists the same certified value as the 2015 record for rigid substrates \cite{Lee2017A}.

From 2015 to 2024, progress centered on tailoring alkali chemistry and refining the absorber microstructure. Solar Frontier’s KF-PDT improved grain-boundary passivation and carrier collection, yielding \textbf{22.3 \%} efficiency in 2015 \cite{Kamada2016New}. Replacing K with heavier alkalis Rb and Cs formed benign, wide-band-gap surface phases (RbInSe$_2$, CsInSe$_2$) whose large ionic radii relieved grain-boundary strain and lowered the diode ideality factor. Partially substituting Cu with Ag further widened the processing window by suppressing copper vacancies and expanding the lattice. With these refinements, Keller~\emph{et al.} pushed the certified efficiency record to \textbf{23.6 \%} by 2023 \cite{keller2024high}. The uninterrupted rise of laboratory performance is plotted in Figure~\ref{fig:cigseff}, which compiles certified results from 2015 to 2024 \cite{kato2018record,nakamura2019cd}.

Looking ahead, modelling indicates that single-junction CIGS could approach its radiative limit of ~25 \% by combining 1.3$-$1.4 eV band-gap absorbers with rear dielectric passivation
(Mo/Al$_2$O$_3$ nanolayers) and dilute-Ag surface layers to further mitigate the \(V_\mathrm{OC}\) deficit. In parallel, monolithic perovskite/CIGS tandems have already surpassed 24 \% efficiency, pointing toward 28$-$30 \% module performance once interconnection and spectral-management losses are minimized.

\subsection{CdTe}

\begin{figure}[!ht]
    \centering
    \includegraphics[width=\columnwidth]{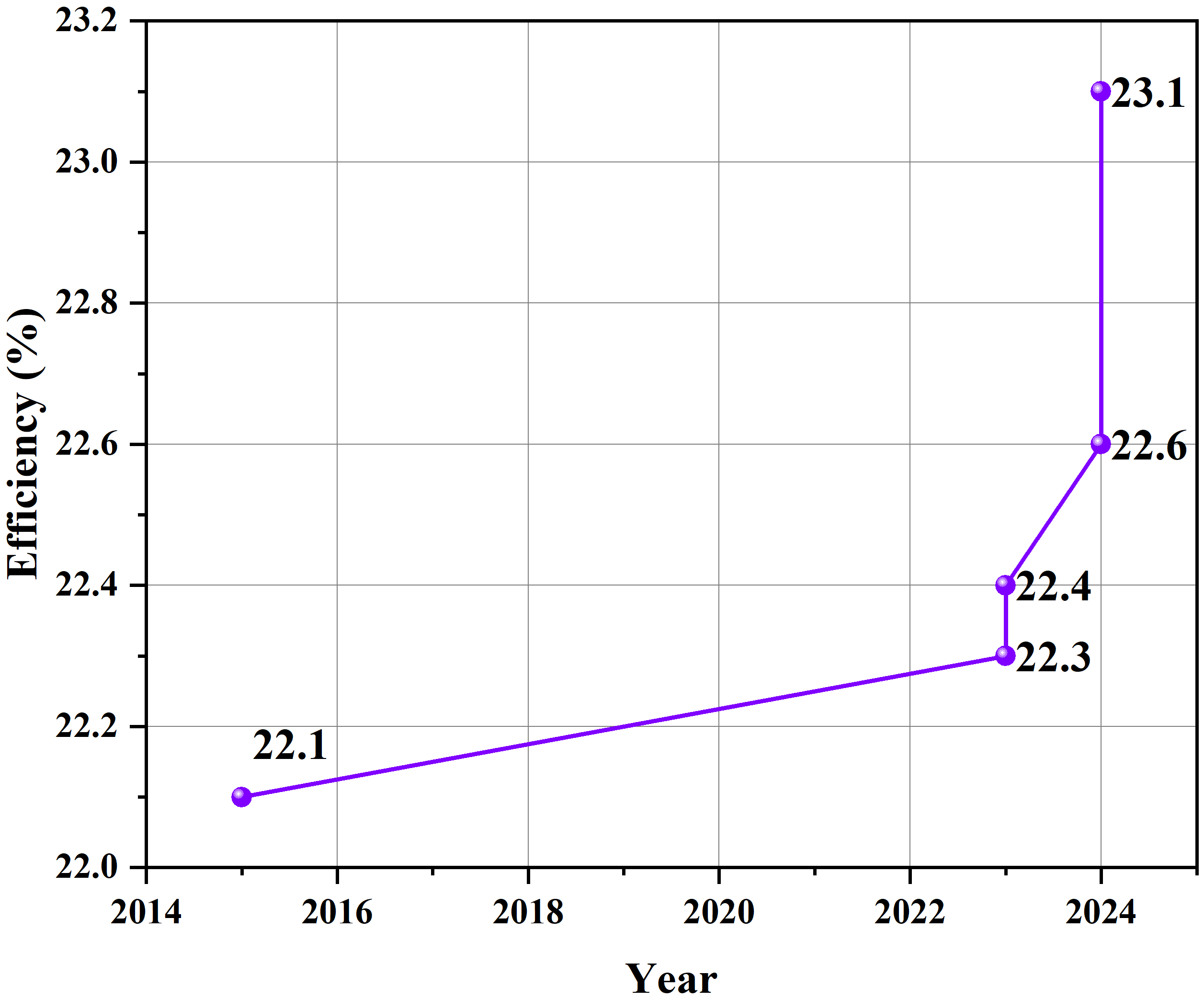}
    \caption{Best laboratory efficiencies for CdTe from 2015 to 2024.}
    \label{fig:cdeteeff}
\end{figure}

Cadmium$-$telluride (CdTe) is a direct‐gap semiconductor with
\(E_g = 1.45\ \mathrm{eV}\) and an absorption coefficient exceeding
\(10^5\ \mathrm{cm}^{-1}\) for \(\lambda < 820\ \mathrm{nm}\).
Consequently, a 1$-$2 $\mu$m film absorbs \(>\!90\%\) of incident AM1.5 photons, enabling low$-$cost, glass$-$integrated devices.
Development began in the early 1970s, when thermally evaporated
CdS/CdTe heterojunctions on SnO\(_2\)$-$coated glass yielded
\textbf{6 \%} efficiency \cite{bonnet1972new}.
Although these first cells were limited by
(i) sub-micron CdTe grains with high twin density and
(ii) an ohmic but recombination$-$active graphite back contact,
they established the now standard \emph{superstrate} architecture
(Glass | TCO | CdS | CdTe | Back contact).

Bube and Kuribayashi revealed that annealing the as‐deposited stack in
CdCl\(_2\) vapour near \(400\,^{\circ}\mathrm{C}\)
recrystallizes CdTe into 3$-$5 $\mu$m grains, heals Te$-$antisite defects, and introduces Cl donors that drive the junction slightly \(n\)-type on the CdTe side, sharpening the depletion width
\cite{mitchell1977evaluation,kuribayashi1983preparation}.
Combined with hotter CdTe growth (close‐space sublimation, CSS) and a
graphite/Ag back electrode, these steps produced \textbf{12.8 \%}
record efficiency by 1983.

Meyers introduced an \emph{$n$$-$$i$$-$$p$} stack thin \(n\)-CdS,
undoped (\(i\)) CdTe, and Cu$-$doped \(p\)-CdTe, achieving more than
\textbf{10 \%} efficiency in 1987 \cite{meyers1988design}.
Cu diffused from the back contact formed shallow acceptors (\(p \sim 10^{14}\ \mathrm{cm}^{-3}\)), but its metastability later emerged as a reliability concern.

\begin{figure}[!ht]
    \centering
    \includegraphics[width=\columnwidth]{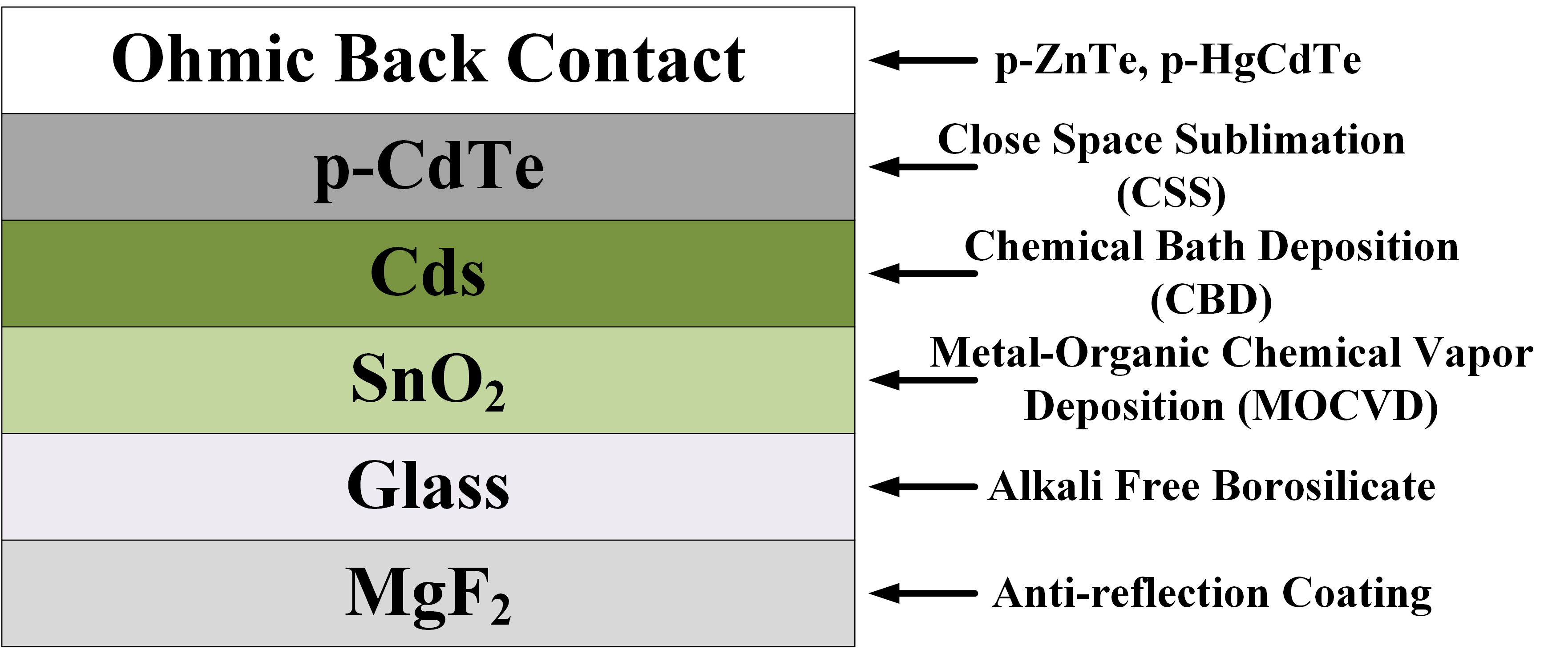}
    \caption{Schematic representation of CdTe/CdS solar cells fabricated using a superstrate configuration \cite{britt1993thin}.}
    \label{fig:CdTe_Early}
\end{figure}

Chu used sputtered SnO\(_2\):F with controlled oxygen vacancies to raise
the front$-$contact work function, while simultaneously optimizing the CdS thickness (60$-$80 nm) to balance transmission and shunt resistance. Precise control of the CdS/CdTe interdiffusion during the CdCl\(_2\) treatment suppressed the so-called “S$-$curve’’ in \(J\)$-$\(V\) behaviour, leading to \textbf{14.6 \%} efficiency in 1992 \cite{chu199113}. Britt’s group then inserted a lightly Cu-doped ZnTe back-surface field (BSF) and replaced the graphite electrode with Mo/Sb\(_2\)Te\(_3\), cutting back-contact recombination velocity below \(10^3\ \mathrm{cm\,s^{-1}}\) and pushing the record to \textbf{15.8 \%} by 1993 \cite{britt1993thin}. Figure~\ref{fig:CdTe_Early} sketches this superstrate structure.

Wu \emph{et al.} at NREL employed high-conductivity,
high‐transparency SnO\(_2\):F (sheet resistance
\(<8\ \Omega_{\square}\)) and replaced CdS with a wider‐gap\, yet lattice matched CdZnS window, trimming parasitic blue loss.
A sputtered ZnTe:Cu/Mo bilayer provided a stable, low-resistance back
contact, enabling \textbf{16.5 \%} efficiency by 2004
\cite{wu200116,wu2004high}.

First Solar industrialized a modified close-space sublimation (CSS) process often described as vapour transport deposition (VTD) that deposits CdTe in <40 s with a steep vertical temperature gradient,
yielding millimetre-scale grains and exceptionally sharp CdS(CdZnS)/CdTe heterointerfaces. In 2015 the company certified a \textbf{22.1 \%} cell by combining:

\begin{enumerate}[label=(\roman*)]
\item CdSe$_x$Te$_{1-x}$ band-gap grading near the junction to mitigate the conduction-band spike,
\item a two-step CdCl$_2$ activation to optimize both grain boundary passivation and Cu diffusion profiles, and
\item an Sb$_2$Te$_3$ back reflector that boosts infrared response \cite{nrel2023}.
\end{enumerate}

Cu$-$based acceptors limit long-term stability because Cu ions out-diffuse under forward bias and temperature. Switching to group-V dopants, most notably arsenic forms deeper, less mobile acceptors (As\(_\mathrm{Te}\)) and lifts the bulk hole density above \(10^{16}\ \mathrm{cm^{-3}}\) without inducing metastability.
With As doping, a refined CdCl\(_2\) activation, and transparent
Mg\(_x\)Zn\(_{1-x}\)O window layers, First Solar reported
\textbf{23.1 \%} certified efficiency in 2024
\cite{mallick2023arsenic,dolia2024four}. Figure~\ref{fig:cdeteeff} charts this steady climb in record performance, confirming CdTe’s status as the highest-efficiency thin-film photovoltaic technology today.

%% file: Emerging.tex
\section{Emerging Thin Film Technologies}

\subsection{Perovskite}

\begin{figure}[!ht]
    \centering
    \includegraphics[width=\columnwidth]{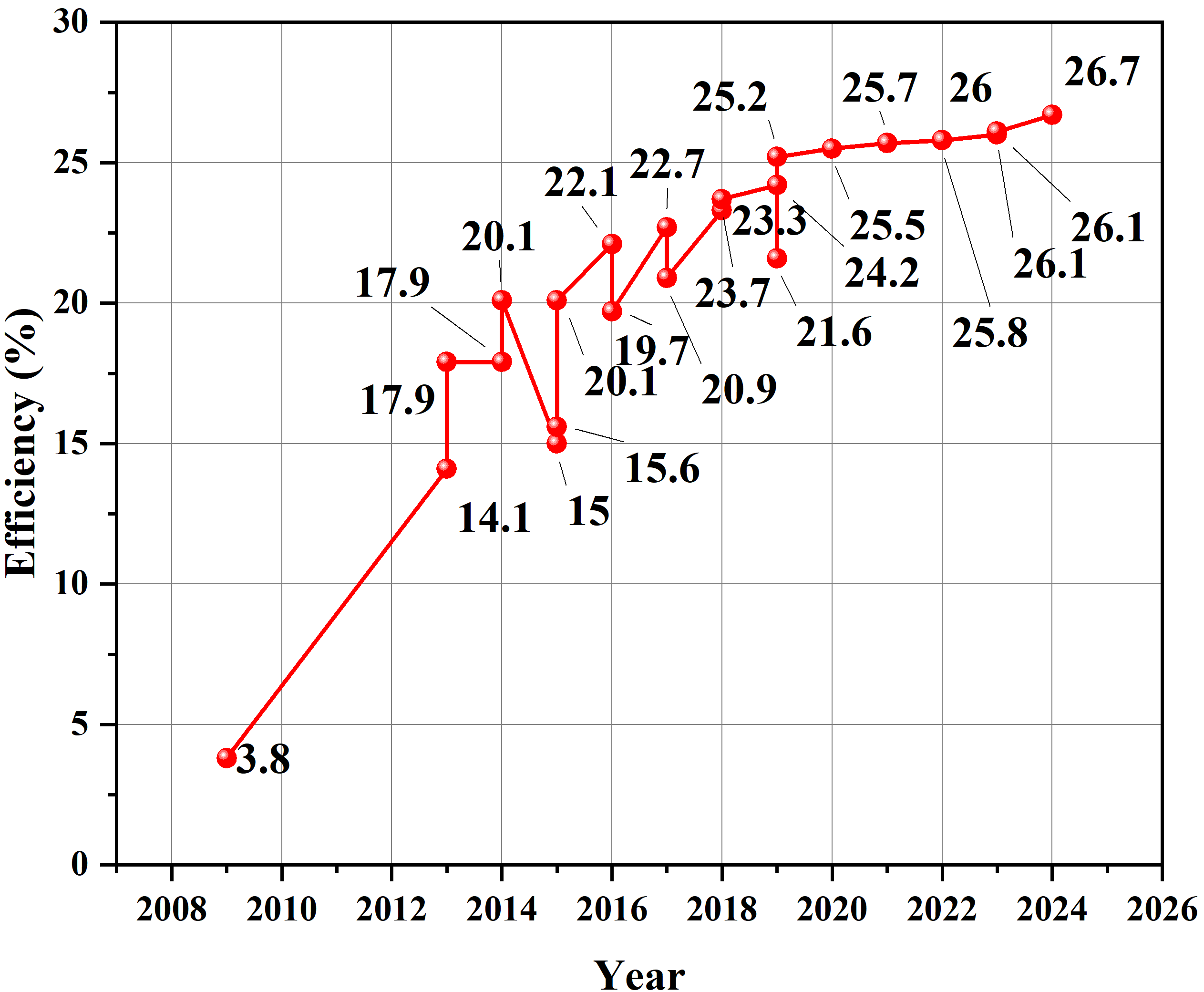}
    \caption{Efficiency improvement of perovskite solar cell}
    \label{fig:peroveff}
\end{figure}

Perovskite solar cells (PSCs) have moved from academic curiosity to front-runner thin-film technology in barely a decade.  The ABX$_3$ crystal family supplies a direct band-gap of 1.45$-$1.65 eV, an absorption coefficient $\alpha > 10^{5}\,\mathrm{cm^{-1}}$ across the visible spectrum, and exciton binding energies below 25 meV, an unusually favorable combination for high-efficiency photovoltaics.  Figure \ref{fig:peroveff} traces the steep, year-on-year rise in certified power-conversion efficiency (PCE).

The modern story begins in 2009, when Kojima \textit{et al.} deployed CH$_3$NH$_3$PbBr$_3$ and CH$_3$NH$_3$PbI$_3$ nanocrystals as photosensitizers in dye-sensitized cells and achieved a modest \textbf{3.8 \%} PCE \cite{kojima2009organometal}.  These liquid-junction devices degraded quickly because the iodide/triiodide electrolyte dissolved the perovskite, yet they revealed exceptional optical absorption and carrier-diffusion lengths exceeding 1 $\mu$m even in polycrystalline films.

Replacing the electrolyte with a solid hole conductor (Spiro-OMeTAD) in 2012 yielded the first all-solid PSCs.  Sequential deposition and vacuum-assisted one-step methods produced pin-hole-free CH$_3$NH$_3$PbI$_{3(1-x)}$Cl$_{3x}$ layers with micrometer-scale grains.  Park’s group subsequently refined antisolvent engineering, dripping chlorobenzene onto a dimethylformamide (DMF) wet film during spin-coating to drive rapid, homogeneous nucleation, increasing certified PCE to \textbf{14.1 \%} in 2013 \cite{park2013organometal}.  Later that year, researchers at KRICT balanced DMF and dimethyl-sulfoxide (DMSO) in the precursor to form a PbI$_2\!\cdot\!\text{DMSO}$ adduct which disassembled slowly, producing dense, uniform perovskite layers and pushing efficiency to \textbf{17.9 \%} \cite{jeon2014solvent}.

Further optimization in 2014 used “toluene drop-casting’’ during the DMSO route to suppress residual PbI$_2$ and cut deep-trap densities below $10^{15}\,\mathrm{cm^{-3}}$, enabling planar CH$_3$NH$_3$PbI$_3$ devices to exceed \textbf{20.1 \%} \cite{yang2015high}.  Parallel work on mesoscopic TiO$_2$/Al$_2$O$_3$ scaffolds showed that well-infiltrated perovskite mitigates hysteresis by allowing ion redistribution under bias.  Quantum-dot PSCs followed in 2015: CsPbI$_3$–FAPbI$_3$ bilayers combined the all-inorganic phase stability of CsPbI$_3$ with the superior optoelectronics of FA-rich perovskite and reached \textbf{15.6 \%} PCE \cite{li2019perovskite}.  Ion-diffusion simulations and Kelvin-probe mapping linked macroscopic hysteresis to vacancy-mediated halide migration, enabling 15 \% stable devices \cite{haruyama2015first}.

Formamidinium (FA) cations fit the A-site more snugly than methylammonium, lowering the Goldschmidt tolerance factor and reducing octahedral tilt.  Mixed-cation FA$_{0.83}$Cs$_{0.17}$Pb(I$_{0.83}$Br$_{0.17}$)$_3$ films fabricated by a one-step antisolvent route delivered \textbf{22.1 \%} certified PCE in 2016 \cite{yang2017iodide}.  Lewis-base additives such as thiocyanate and DMSO passivated Pb-rich facets, while phenethylammonium iodide formed a two-dimensional Ruddlesden$-$Popper (2D/3D) capping layer that sealed grain boundaries against moisture.  Surface fluorination combined with alkyl-ammonium bromide treatments pushed planar-SnO$_2$ devices to \textbf{23.7 \%} by 2018 \cite{jiang2019surface,jiang2017planar}.

Isoelectronic Mg$^{2+}$ and Sr$^{2+}$ doping on the Pb site lowered the formation energies of vacancy complexes and cut non-radiative recombination losses.  Jung \textit{et al.} optimized the Br/I ratio to flatten the conduction-band offset at the SnO$_2$ interface, achieving \textbf{24.2 \%} in 2019 \cite{jung2019efficient}.  Carrier-selective contacts$-$Me$-$4PACz self-assembled monolayers for holes and PC$_{61}$BM:C$_{60}$ bilayers for electrons raised the external radiative efficiency above 20 \%, culminating in a \textbf{25.2 \%} record \cite{yoo2021efficient}.  Min \textit{et al.} then inserted an atomically coherent FAI/CsI interlayer that relieved lattice strain and suppressed ion migration, reaching \textbf{25.5 \%} in 2020 \cite{min2021perovskite}.

Replacing Spiro-OMeTAD with PTAA or Me-4PACz self-assembled monolayers eliminated hygroscopic LiTFSI dopant and reduced parasitic absorption.  Combined with fullerene-based passivation of SnO$_2$ surface states, devices stabilized at \textbf{25.7 \%} in 2021 and \textbf{25.8 \%} in 2022 \cite{min2021perovskite}.  A resonant cavity formed by a transparent MoO$_x$/Ag/MoO$_x$ rear contact increased the optical path length, while inverted (p-i-n) cells employing carboxylate-anchored SAMs eradicated ion-induced hysteresis to yield a \textbf{26.1 \%} certified world record in 2023 \cite{feng2023resonant,li2024revolutionary}.

Halide-segregation control delivered the next milestone.  Meng \textit{et al.} found that trace ZnCl$_2$ generates a quasi-2D ZnI$_2$ passivation layer that pins halides, suppressing photo-induced Br/I segregation and widening the quasi-Fermi-level splitting.  Together with guanidinium-anchored halide scavengers and vacuum-flashed antisolvent deposition, they achieved an impressive \textbf{26.7 \%} PCE in 2024 \cite{meng2024inhibition}.  Stability tests under 1-sun, 65 \textdegree C and 85 \% RH showed under 5 \% performance loss after 1 000 h, underscoring the commercial promise of perovskite photovoltaics.


\subsection{CZTS}

\begin{figure}[!ht]
    \centering
    \includegraphics[width=\columnwidth]{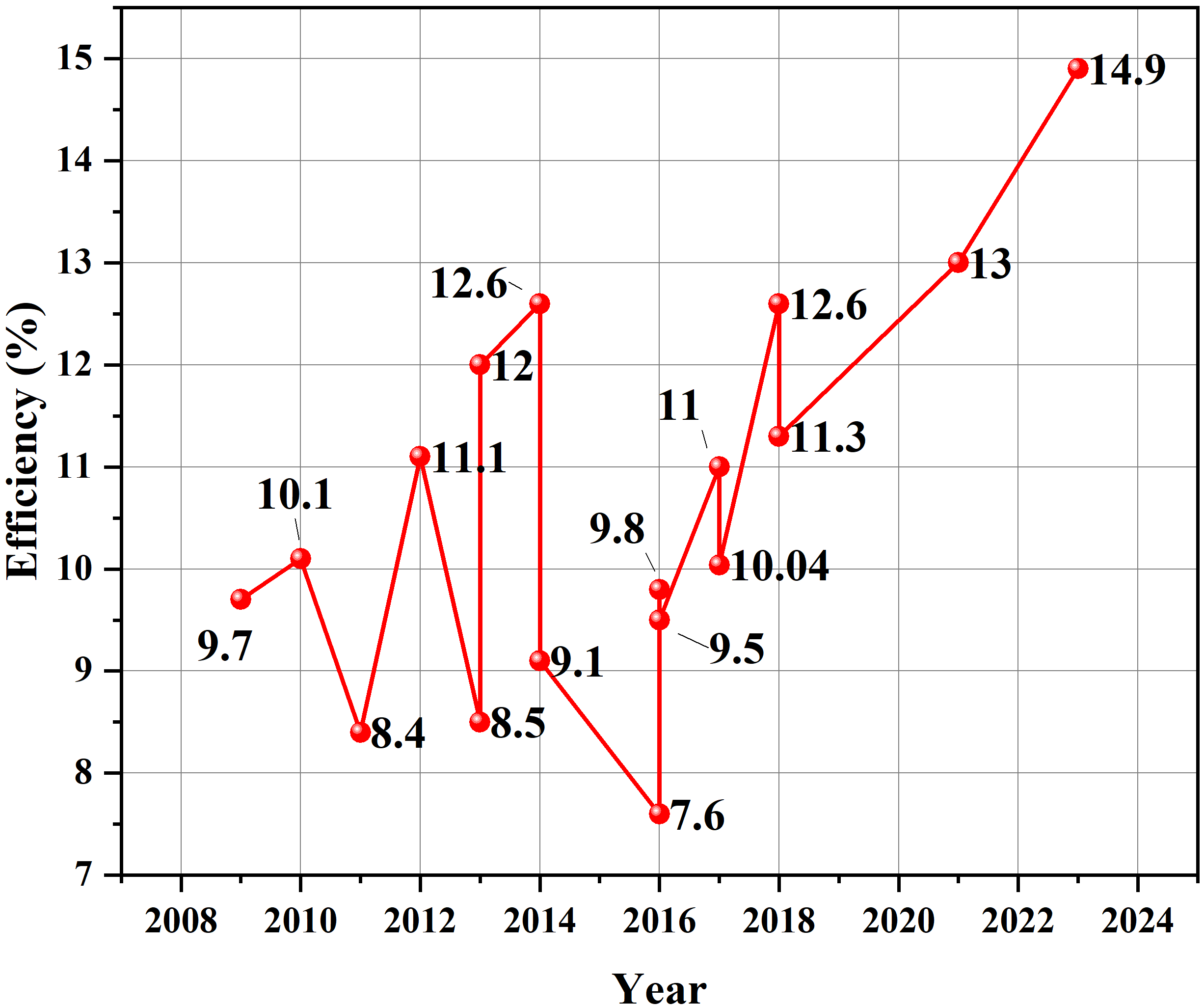}
    \caption{Efficiency improvement of CZTS cell}
    \label{fig:czts}
\end{figure}

Cu$_2$ZnSnS$_4$ (CZTS) has attracted sustained interest as a
Cd- and In-free thin-film absorber because its constituent elements are both earth-abundant and non-toxic.  In the kesterite$-$stannite lattice (the two polytypes differ only by cation ordering) the material exhibits a direct band-gap of 1.4$-$1.6 eV and an absorption coefficient exceeding $10^{4}\,\mathrm{cm^{-1}}$ for $h\nu\!>\!1.6$ eV, so a 1$-$2 $\mu$m film can, in principle, harvest most of the terrestrial solar spectrum.  The chief obstacles are severe Cu$-$Zn cation disorder, their ionic radii differ by
just 0.06 \AA, and deep Sn$_\mathrm{Zn}$ antisites, which together impose a pronounced open-circuit-voltage (\(V_\mathrm{OC}\)) deficit relative to the band-gap energy.

\begin{figure}[!ht]
   \centering
    \includegraphics[width=.9\columnwidth]{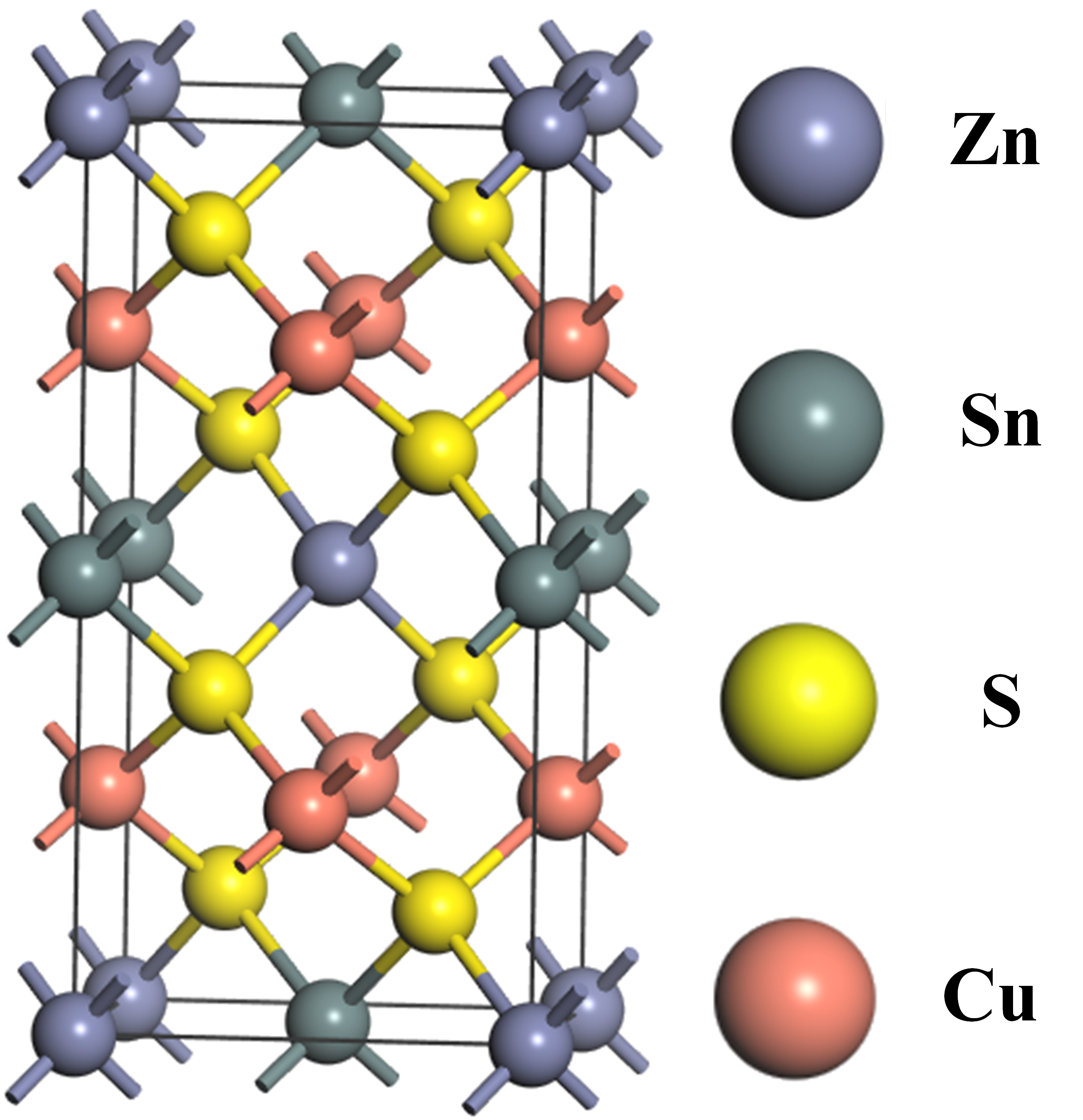}
    \caption{Stannite crystal structure for CZTS}
    \label{fig:stannite}
\end{figure}

The first proof-of-concept device, reported in 1996, was fabricated by electron-beam evaporation of Cu, Zn and Sn layers followed by
sulphuration at $520\,^{\circ}\mathrm{C}$; the resulting stannite structure (Figure \ref{fig:stannite}) contained sub-micrometer grains and numerous Cu$_2$SnS$_3$ inclusions, limiting the power-conversion efficiency (PCE) to \textbf{0.66 \%} \cite{katagiri1997preparation}.  Replacing the high-resistance undoped ZnO window with Al-doped ZnO and inserting a
thin CdS buffer to mitigate interface recombination raised the PCE to
\textbf{2.3 \%} the following year \cite{friedlmeier1997heterojunctions}.  A decade of vacuum-process optimization maintaining Cu/(Zn+Sn)\,$\approx$ 0.8$-$0.9 to stay Cu-poor and supplying Na from soda-lime glass pushed the record to
\textbf{5.7 \%} in 2005 \cite{jimbo2007cu2znsns4}.

Figure \ref{fig:czts} summarizes these early milestones and the rapid
progress that followed.  Solution processing reshaped the field in 2009: a hydrazine-based molecular ink produced centimeter-scale grains, a compact morphology, and, crucially suppressed Sn loss during high-temperature annealing, yielding a \textbf{9.7 \%} CZTS device \cite{todorov2010high}.  Alloying the anion lattice with selenium in 2010 lowered the band-gap to $\approx$1.1 eV, extended minority-carrier diffusion lengths, and delivered the first
Cu$_2$ZnSn(S,Se)$_4$ (CZTSSe) cell to exceed \textbf{10.1 \%}
\cite{barkhouse2012device}.  Vacuum thermal evaporation achieved \textbf{8.4 \%} in 2011, while refinements of the hydrazine route, using Cu/(Zn+Sn)\,= 0.75 absorbers pushed CZTS$_{0.3}$Se$_{0.7}$ efficiency to \textbf{11.1 \%} in 2012 \cite{todorov2013beyond}.  A simpler thiourea/thioglycolic-acid ink, compatible with roll-to-roll coating, still reached \textbf{8.5 \%} \cite{guo2012simple}.

Annealing in precisely tuned S/Se vapor in 2014 created a graded band-gap ($\approx$1.0 eV at the Mo back contact, 1.15 eV near the CdS interface), reduced interface recombination, and set a CZTSSe record of \textbf{12.6 \%} \cite{wang2014device}.  Lower-temperature H$_2$S treatments curtailed SnS$_x$ evaporation and lifted the pure-sulphide CZTS efficiency to \textbf{9.1 \%} \cite{fukano2013enhancement}. Continued fine-tuning of cation and anion ratios produced \textbf{7.6 \%} and \textbf{9.5 \%} devices in 2016 \cite{sun2016over};
by 2017, meticulous control of Cu deficiency and the Zn/Sn balance
raised the all-sulphide record to \textbf{10.0 \%}, while mixed-anion
CZTSSe peaked at \textbf{11.3 \%}
\cite{yan2018cu2znsns4,son2019effect}.

A hydrazine-free chloride-complex ink, combined with (Ag,Na) alkali
doping to passivate grain boundaries, enabled a certified
\textbf{13.0 \%} CZTSSe cell in 2021 \cite{gong2022elemental}.  The
current record, \textbf{14.9 \%} (2023), was achieved by engineering a Cu-poor, Zn-rich surface during selenization; this suppressed deep
Sn$_\mathrm{Zn}$ defects, narrowed the band-tailing energy
($E_\mathrm{U}<\!30$ meV), and lifted \(V_\mathrm{OC}\) above 650 mV
\cite{li2024suppressing}.

Despite this progress, kesterite devices remain well below their
$\approx$32 \% Shockley$-$Queisser limit because the \(V_\mathrm{OC}\) deficit still exceeds 0.40 V.  Current research therefore targets (i) partial substitution of Cu with Ag or Cd to alleviate Cu$-$Zn disorder; (ii) Ge or Sb alloying to passivate Sn-related traps; (iii) benign Zn(O,S) or In$_2$S$_3$ buffer layers to eliminate the conduction-band cliff at the CdS/CZTS(e) interface; and
(iv) rear dielectric passivation to curtail MoSe$_2$ back-contact
recombination.  Coupled with advanced light-management schemes, these
measures could lift single-junction efficiencies past 18 \%, positioning kesterite as a viable bottom cell in tandem stacks with silicon or perovskite top absorbers.

\subsection{Quantum Dot}

\begin{figure}[!ht]
    \centering
    \includegraphics[width=\columnwidth]{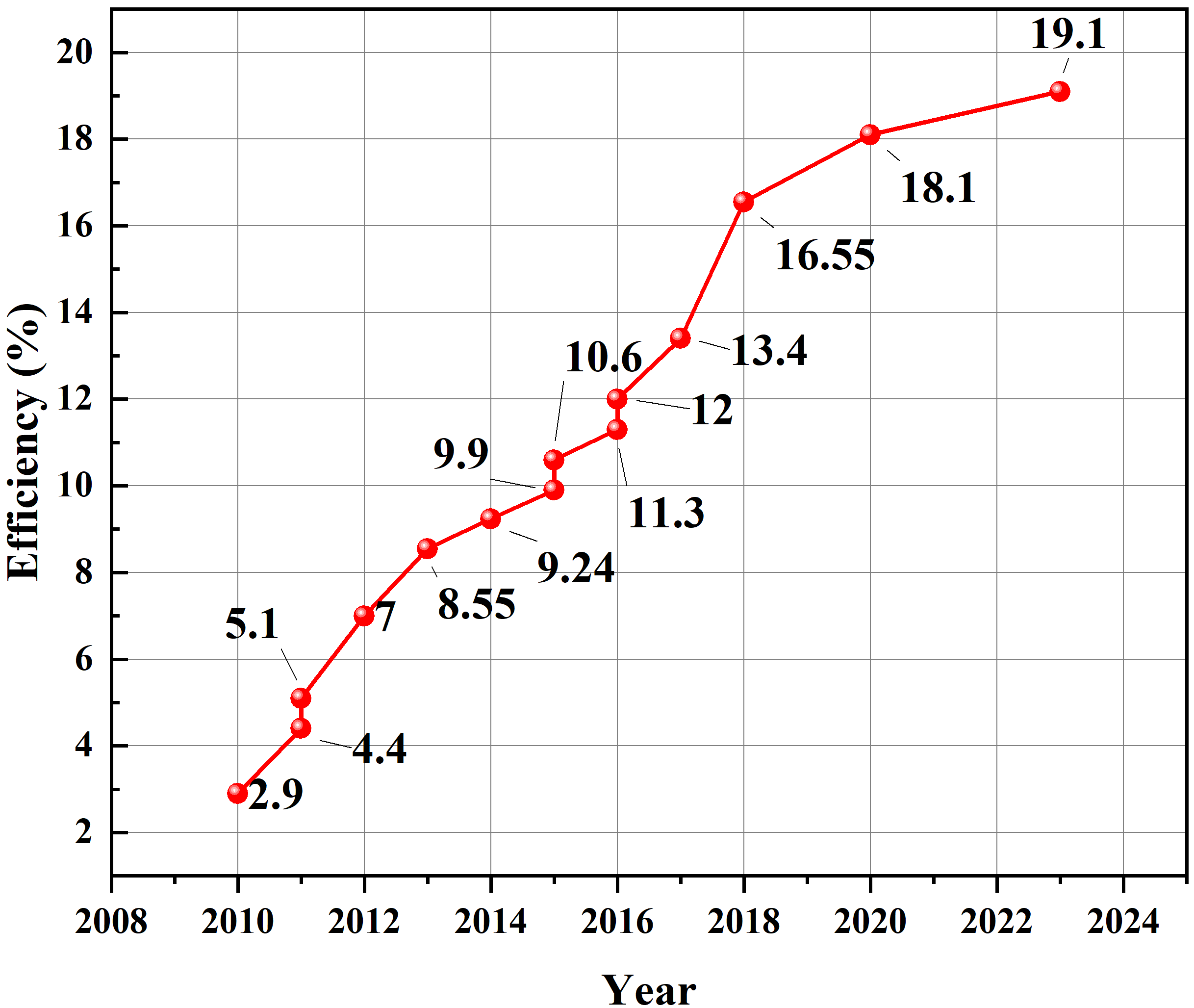}
    \caption{Efficiency improvement of QD cell}
    \label{fig:qdeff}
\end{figure}

Quantum-dot (QD) photovoltaics exploit quantum confinement effects in colloidal nanocrystals whose band gap \(E_g\) can be tuned over 1.0$-$2.0 eV simply by varying particle size or composition, making them attractive for single-junction optimization and multi-junction designs. Solution processability, low thermal budgets (\(<150^{\circ}\mathrm{C}\)), and compatibility with flexible substrates further position QD cells as a scalable alternative to vacuum-deposited CdTe and CIGS.  The rapid efficiency climb summarized in Figure \ref{fig:qdeff}, reflects steady progress in surface passivation, carrier-transport layers, and device architecture.

\begin{figure}[!ht]
    \centering
    \includegraphics[width=\columnwidth]{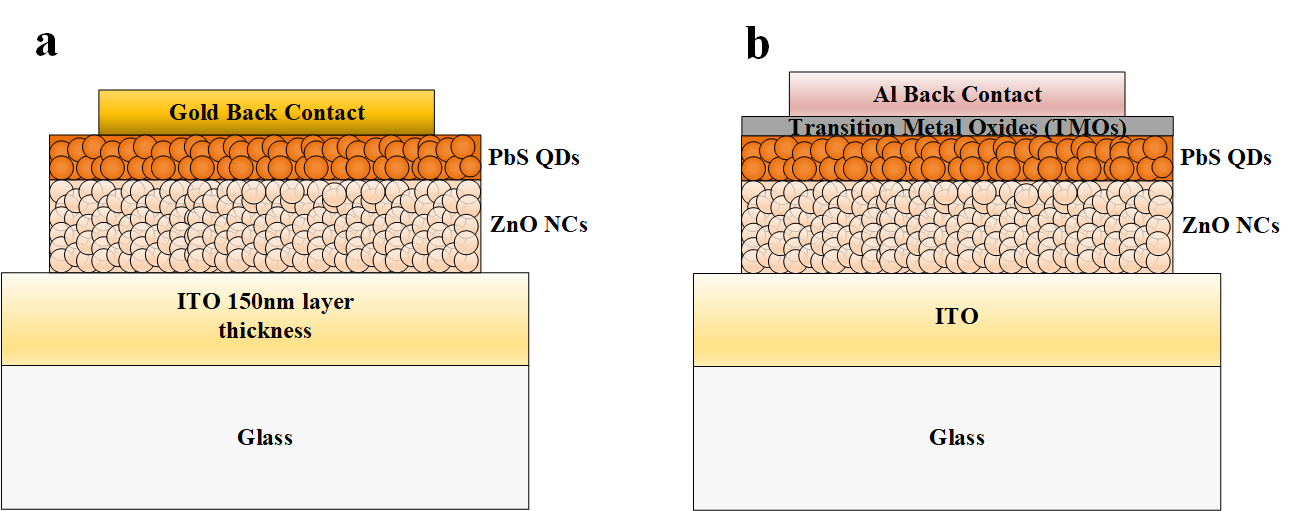}
    \caption{(a) Structure of QD cell propose by NREL \cite{luther2010stability} (b) Imrpoved QD structure by introducing TMOs layer between the Al back contact and PbS QD \cite{gao2011n}}
    \label{fig:qstructure}
\end{figure}

The first certified device, reported by NREL in 2010, combined air-stable PbS QDs passivated with 1,2-ethanedithiol (EDT) and a ZnO nanoparticle electron-selective contact in an inverted
ITO|ZnO|PbS-QD|MoO\(_x\)|Ag stack (Figure \ref{fig:qstructure}),
achieving \textbf{2.9 \%} efficiency and retaining 95 \% of its initial output after 1000 hrs of one-sun operation
\cite{luther2010stability}.  In 2011 the same group replaced the organic MoO\(_x\) hole-extraction layer with transition-metal oxides
(MoO\(_x\), V\(_2\)O\(_x\)) deposited by thermal evaporation, raising
the work-function alignment and boosting \(V_\mathrm{OC}\) to 0.45 V,
which lifted the PCE to \textbf{4.4 \%} \cite{gao2011n}.  Concurrently, the University of Toronto shortened the
QD ligands from oleic acid to butylamine/EDT and optimised particle size (\(E_g\approx1.3\ \mathrm{eV}\)); improved electronic coupling reduced series resistance and enabled a \textbf{5.1 \%} depleted-heterojunction device \cite{pattantyus2010depleted}.

Toronto researchers next adopted a layer-by-layer \- spin/spray process in which each 40$-$50 nm QD monolayer underwent solution ligand-exchange with 1,2-ethanedithiol, yielding crack-free films up to 400 nm thick and a certified \textbf{7.0 \%} PCE in 2012
\cite{ip2012hybrid}.  MIT then introduced hybrid passivation: a two-step exchange that first substituted oleate with tetrabutyl-ammonium iodide (TBAI) and subsequently grafted EDT, reducing surface trap densities to \(2\times10^{15}\ \mathrm{cm^{-3}}\).  Longer carrier lifetimes (>$300$ ns) pushed the 2013 record to \textbf{8.55 \%}
\cite{chuang2014improved}.  Finer control of iodide/thiolate ratios in 2014 raised \(V_\mathrm{OC}\) above 0.6 V and set a \textbf{9.24 \%} mark \cite{lan2016passivation}; molecular-iodine oxidative cleansing further removed sub-bandgap states, reaching \textbf{9.9 \%} early in 2015. A solvent-polarity-engineered methylammonium-iodide (MAI) halide-passivation step then delivered the first double-digit result, \textbf{10.6 \%}, by mitigating surface recombination and enhancing hole extraction \cite{lan201610}.

In 2016, solution-phase ligand-exchanged “CQD inks’’ enabled blade- and slot-die coating of dense films with an order-of-magnitude higher
carrier mobility ($\approx$1 cm\(^2\) V\(^{-1}\) s\(^{-1}\)), advancing the record to \textbf{11.3 \%} \cite{liu2017hybrid}.  Subsequent incorporation of two-dimensional PbI\(_2\) sheets as an inorganic passivant at grain boundaries cut the tail-state Urbach energy below 25 meV and produced \textbf{12.0 \%} by year’s end \cite{xu20182d}. NREL’s switch to CsPbI\(_3\) perovskite QDs in a similar inverted stack yielded a \textbf{13.4 \%} cell with a remarkably low voltage deficit (\(E_g/q-V_\mathrm{OC}=0.37\) V), underscoring the potential of lead-halide QDs for tandem applications \cite{christians2018perovskite}.

A ligand-assisted cation-exchange (LACE) scheme developed at the
University of Queensland in 2018 replaced Cs\(^{+}\) partially with
FA\(^{+}\)/MA\(^{+}\), stabilizing the black perovskite phase and
raising the certified PCE to \textbf{16.55 \%} \cite{hao2020ligand}.  UNIST researchers introduced butylhydroxytoluene (BHT) as a radical scavenger and cross-linker in 2020; BHT suppressed halide migration and improved film cohesion, pushing efficiency to \textbf{18.1 \%}
\cite{aqoma2024alkyl}.  Continued optimization of surface dipoles,
energy-level alignment, and graded halide distribution culminated in a \textbf{19.1 \%} certified world record for QD photovoltaics in 2023 \cite{nrel2023}.

Key challenges ahead include reducing non-radiative recombination below the radiative limit (~10 mA cm\(^{-2}\)), engineering robust
charge-selective contacts that withstand >1 000 h under >85 $^\circ$C/85 \% RH, and scaling CQD inks to meter-scale roll-to-roll lines without film cracking.  Addressing these issues could soon push single-junction QD cells beyond the 20 \% threshold and enable high-voltage perovskite$-$quantum-dot tandem architectures.

\subsection{Organic Photovoltaics}

\begin{figure}[!ht]
    \centering
    \includegraphics[width=\columnwidth]{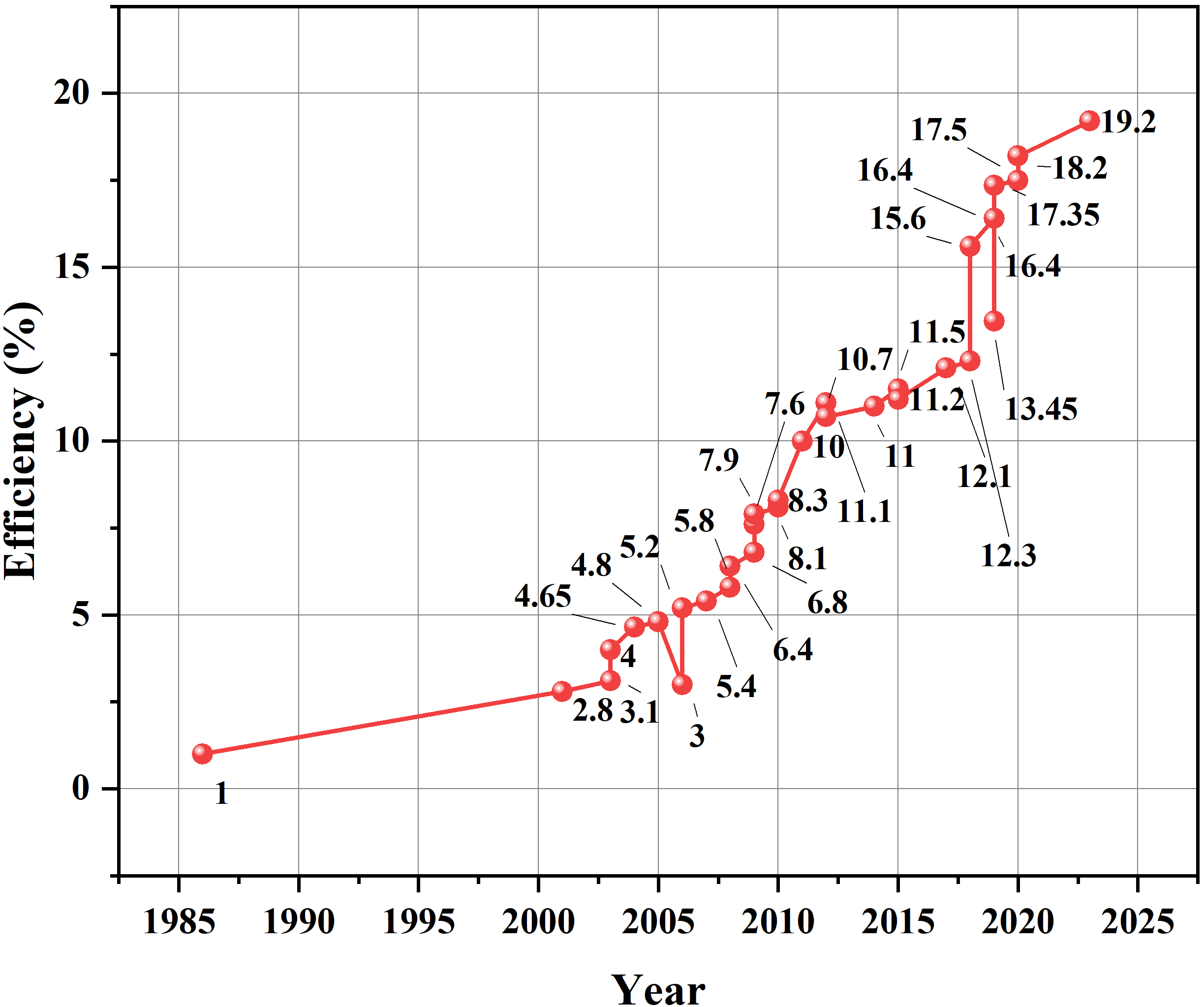}
    \caption{Progression of OPV cell}
    \label{fig:opv}
\end{figure}

\begin{figure}[!ht]
    \centering
    \includegraphics[width=.9\columnwidth]{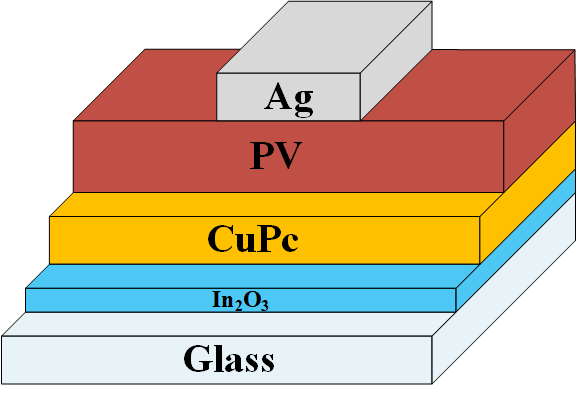}
    \caption{Configuration of a two-layer organic cell: ITO-coated glass , 300 $\si{\angstrom}$ copper phthalocyanine (CuPc) layer, 500 $\si{\angstrom}$ pyrylene tetracarboxylic derivative (PV) layer, and an opaque Ag layer, all deposited by vacuum evaporation \cite{tang1986two}}
    \label{fig:twolayers}
\end{figure}

Organic photovoltaics (OPVs) convert light into electricity by photo-induced charge transfer between a donor and an acceptor whose highest-occupied and lowest-unoccupied molecular orbitals (HOMO$-$LUMO) differ by more than the $\approx$0.3 eV exciton-binding energy typical of organic semiconductors. Because singlet-exciton diffusion lengths are only 5$-$20 nm, modern OPVs employ bulk heterojunctions (BHJs) in which the donor and acceptor interpenetrate on this length scale, creating a bicontinuous network for charge separation and transport. The field began with C.W.Tang’s bilayer cell at Eastman Kodak, which paired copper-phthalocyanine (CuPc) with a perylene tetracarboxylic derivative; that two-layer device (Figure \ref{fig:twolayers}) delivered a then-remarkable \textbf{1 \%} efficiency and proved that all-organic materials could generate photocurrent \cite{tang1986two}. Figure \ref{fig:opv} charts the subsequent rise in certified power-conversion efficiency from Tang’s pioneering result to today’s near-20 \% records, and sketches the key device architectures that enabled each milestone, serving as a visual roadmap of OPV progress.

The first major leap came when conjugated polymers were blended with soluble fullerene acceptors (PC\textsubscript{60}BM), increasing the donor$-$acceptor interfacial area by orders of magnitude. Optimizing spin-coating to curb micron-scale phase segregation pushed efficiencies to \textbf{2.8 \%}, \textbf{3.1 \%}, and ultimately \textbf{4.0 \%} in early polymer:fullerene systems \cite{shaheen20012,nrel2023}. Incorporating regioregular P3HT and controlling solvent drying increased crystallinity, raising the figure to \textbf{4.65 \%}.

Thermal annealing of P3HT:PCBM at Konarka induced edge-on P3HT stacking and raised hole mobility, producing \textbf{4.8 \%} efficiency \cite{ma2005thermally}. Adding 1,8-octanedithiol as a crystallization additive promoted fibril formation and yielded \textbf{5.2 \%} \cite{peet2007efficiency}. Plextronics’ pre-formulated Plexcore inks enabled printed modules at \textbf{5.4 \%} and \textbf{5.8 \%} \cite{laird2007advances}. Low-band-gap polymers such as PTB and PBDT extended absorption to 750 nm; careful control of nanoscale morphology lifted laboratory efficiencies to \textbf{7.9 \%} \cite{brabec2010polymer,chen2009polymer}. Benzothiadiazole-imide acceptors with higher electron affinity reduced voltage losses, and solution-processed devices reached \textbf{8.3 \%} \cite{bloking2011solution}.

Donor$-$acceptor copolymers like PTB7-Th, paired with fullerene derivatives (PC\textsubscript{71}BM, ICBA), exploited higher dielectric constants and optimized HOMO alignment, surpassing \textbf{11.5 \%} PCE \cite{hu2015terthiophene,mori2015organic}. Further gains stalled until non-fused-ring acceptors (NFAs) such as ITIC and Y6 appeared. Their strong near-infrared absorption, high electron mobility, and shallow LUMO levels cut the voltage deficit below 0.55 V; blade-coated PTB7-Th:ITIC devices exceeded \textbf{12 \%} \cite{luo2019reduced}.

Casting PM6:Y6 films from non-halogenated solvents with a vertical phase-gradient yielded 15$-$25 nm domains and mixed face-on/edge-on orientation, delivering a certified \textbf{16.4 \%} \cite{li2020vertical,sun2019monothiophene}. Sequential “green” processing in anisole preserved this morphology and pushed the single-junction record to \textbf{18.2 \%} \cite{zhu2022single,ma2020promoting,li2019generic}.

A ternary donor$-$acceptor$-$acceptor strategy added crystalline donor polymer D18 to the PM6:L8-BO matrix, forming a double-fibril network that lengthened exciton diffusion to 18 nm, boosted carrier mobility to approximately \(4\times10^{-4}\,\mathrm{cm^{2}\,V^{-1}\,s^{-1}}\), and achieved an 82 \% fill factor. These optimizations produced a certified \textbf{19.2 \%} world record \cite{zhu2022single}. Inverted cells with ZnO and MoO\(_x\) charge-selective layers now retain more than 90 \% of their initial power after 1 000 h at 85 \textdegree C and 85 \% RH, underscoring the technology’s accelerating commercial viability.

Looking ahead, the critical targets are to reduce non-radiative voltage losses below 0.15 V, suppress burn-in degradation via end-group dipole control, and scale slot-die-coated NFA formulations to meter-wide web lines. Tandem stacks that combine a wide-gap PM6:Y14 front cell (1.9 eV) with a narrow-gap D18:L8-BO rear cell (1.3 eV) have already demonstrated 24 \% four-terminal efficiency, suggesting that OPVs could soon rival thin-film inorganics in both performance and environmental footprint.

\subsection{DSSC}

\begin{figure}[!ht]
    \centering
    \includegraphics[width=\columnwidth]{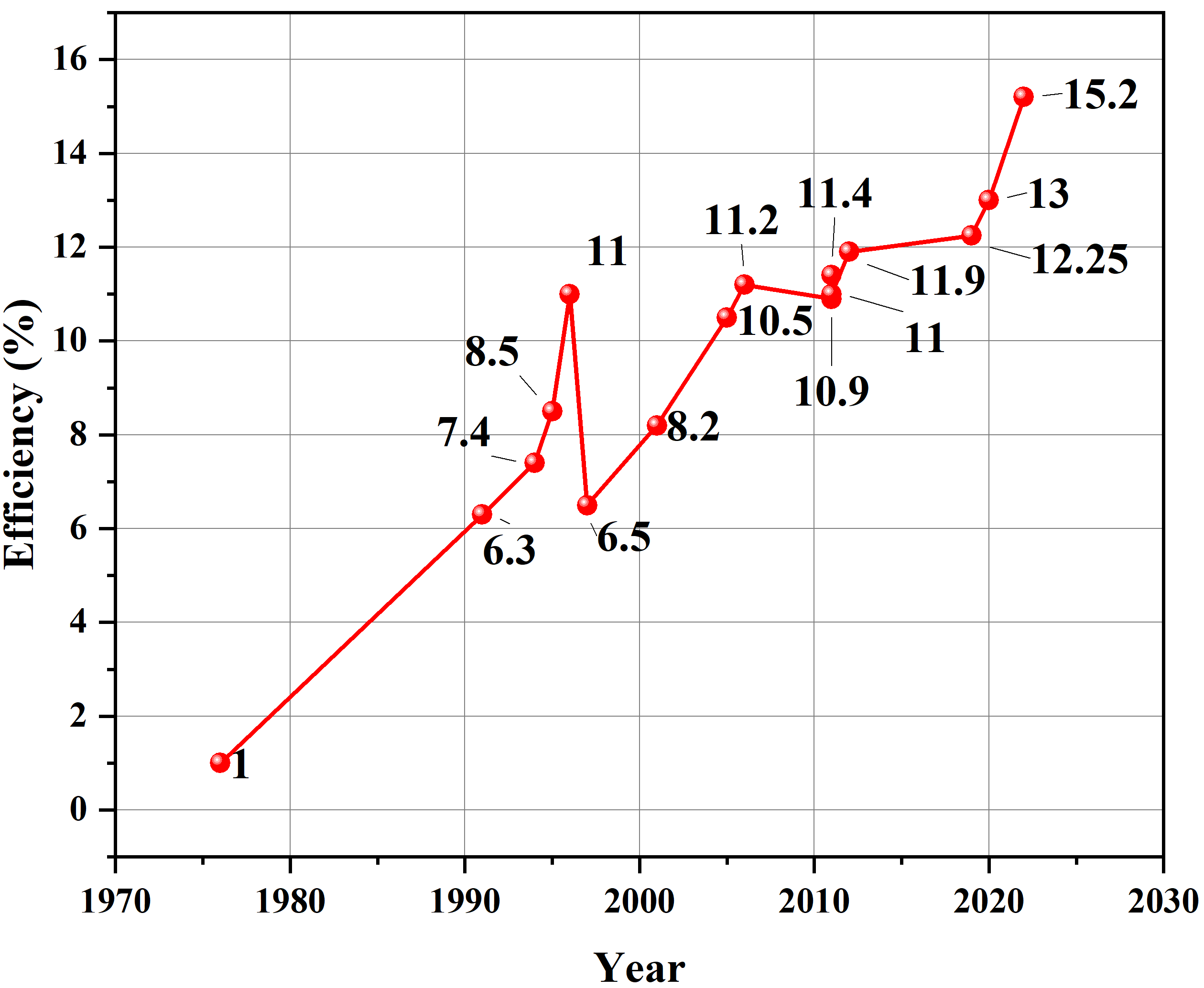}
    \caption{Evolution of DSSC cell}
    \label{fig:dssc}
\end{figure}

Dye-sensitized solar cells (DSSCs) generate photocurrent by mimicking a single photosynthetic step: a light-absorbing dye harvests photons, injects electrons into the conduction band of a wide-gap semiconductor, and a redox shuttle then regenerates the oxidized dye.  Figure \ref{fig:dssc} charts the certified efficiency records of DSSCs, from the earliest 1 \% prototypes to the latest devices exceeding 15 \%,and highlights the key material and architectural innovations that enabled each milestone.  Early demonstrations in the 1970s used micrometer-sized ZnO or TiO\(_2\) particles coated with erythrosine or Rose Bengal and achieved barely \textbf{1 \%} efficiency, largely because the planar particle surfaces limited dye loading and the electrolytes suffered rapid recombination losses \cite{tsubomura1976dye}.

The breakthrough came in 1991 when O'Regan and Grätzel introduced a \textit{mesoporous} anatase TiO\(_2\) electrode composed of 10–20 nm crystallites sintered into a 10$-$12 $\mu$m-thick, 60 \% porous film \cite{gratzel1991low}.  Its internal surface area exceeded 100 m\(^2\) cm\(^{-3}\), enabling monolayer adsorption of the ruthenium bipyridyl dye \ce{[Ru(dcbpy)2(NCS)2]} (N3) and near-quantitative visible-light harvesting.  Combined with an I\(^{-}\)/I\(_3^{-}\) electrolyte and a Pt-coated counter-electrode, the device reached \textbf{6.3 \%} efficiency an order-of-magnitude leap over previous records and established the classic liquid-electrolyte DSSC architecture.

Subsequent optimization focused on film morphology, co-adsorbents and electrolyte additives.  Narrowing the TiO\(_2\) particle-size distribution and adding a 4$-$5 $\mu$m light-scattering over-layer of 400 nm rutile spheres boosted the optical path length, while chenodeoxycholic acid (CDCA) suppressed dye aggregation, pushing efficiency to \textbf{7.4 \%} by 1994 \cite{lampert1994towards}.  Tert-butylpyridine (TBP) shifted the TiO\(_2\) conduction-band edge ~80 mV negative and reduced surface recombination with I\(_3^{-}\), yielding \textbf{8.5 \%} by 1995.  Ligand engineering of Ru dyes (N719, Z907, “black dye”) and ionic-liquid electrolytes curbed solvent volatility, lifting certified performance to \textbf{10.5 \%} in the early 2000s \cite{meyer2001recent,wang2005significant}.

Because liquid electrolytes limited operational lifetime, researchers explored quasi-solid gels and solid-state hole-transport materials (HTMs) such as spiro-OMeTAD.  Although early solid-state devices lagged behind their liquid counterparts, improvements in pore filling and interface energetics preserved efficiencies above 7 \%.  Parallel molecular design yielded donor-$\pi$-acceptor zinc-porphyrin dyes (e.g., YD2-o-C8) whose broad Soret and Q-band absorption covered 350$-$750 nm; co-sensitisation with silyl-anchor organic dyes boosted photocurrent, delivering \textbf{13 \%} efficiency in 2014 \cite{mathew2014dye}.

During the past decade, cobalt-bipyridine and copper-phenanthroline redox couples have replaced I\(^{-}\)/I\(_3^{-}\), offering 200$-$300 mV higher open-circuit voltage while retaining fast dye regeneration.  Tailoring the ligand field around Co\(^{2+/3+}\) optimized diffusion coefficients and suppressed recombination; hydrophobic alkyl spacers on porphyrin donors mitigated electrolyte ingress.  Together with hierarchical TiO\(_2\) photo-anodes and SnO\(_2\):TiO\(_2\) core$-$shell scaffolds, these advances recently pushed certified DSSC efficiency beyond \textbf{15 \%} \cite{ren2023hydroxamic}.  The champion device employs a hydroxamic-acid-anchored porphyrin dye whose LUMO couples strongly to the TiO\(_2\) lattice, a cobalt-bipyridine electrolyte with 0.1 M TBP for optimal band alignment, and a sputtered Pt/PEDOT counter electrode that keeps series resistance below 2 $\Omega$ cm\(^2\).

Solid-state DSSCs now exceed 13 \% with Cu(tmby)\(_2\)/TFSI HTMs, while perovskite$-$dye tandems reach 28 \% in four-terminal mode by stacking a 1.73 eV perovskite top cell over a 1.1 eV porphyrin-sensitised TiO\(_2\) bottom cell.  Current priorities include passivating TiO\(_2\) surface traps with ultrathin Al\(_2\)O\(_3\), replacing volatile solvents with eutectic melts, and developing panchromatic organic sensitizers whose excited-state lifetimes exceed 50 ns on TiO\(_2\).  These efforts aim to narrow the remaining efficiency gap to crystalline silicon while retaining DSSCs’ hallmark low-cost, low-temperature fabrication.

%% file: CommercialModule.tex
\section{Module activities}

\begin{center}
\footnotesize
\captionof{table}{Commercial thin-film module/sub-module efficiencies}
\label{tab:commercial}
\begin{tabular}{l c c p{2.0cm}}
  \hline
  \textbf{Type} & \textbf{Eff. (\%)} & \textbf{Area (cm$^{2}$)} & \textbf{Company} \\
  \hline
  c-Si        & 24.9\cite{maxeon2024benchmark}         & 17753 & Maxeon \\
  CIGS        & 19.2\cite{sugimoto2014high}            & 841   & Solar Frontier \\
  CdTe        & 19.9\cite{firstsolar_series6}          & 23932 & First Solar \\
  Perovskite  & 21.95\cite{solaeon_perovskite_2024}    & 1200  & SolaEon \\
  OPV         & 14.46\cite{pv_magazine2024organic}     & 204   & HI ERN \\
  OPV         & 13.1\cite{pv_magazine2023efficiencies} & 1475  & Waystech/\newline Nano \\
  DSSC        & 8.8\cite{hernandez2011technical}       & 398.8 & Sharp \\
  \hline
\end{tabular}
\end{center}

Table \ref{tab:commercial} compares state-of-the-art commercial \emph{module} efficiencies for the main thin-film technologies CdTe, CIGS, \textit{a}-Si:H/$\mu$c-Si:H tandems, perovskites, organic photovoltaics (OPVs), CZTS(e), quantum-dot (QD) devices, and dye-sensitized solar cells (DSSCs), with today’s best crystalline-silicon (c-Si) modules.  Although mono- and n-type c-Si still post the highest commercial efficiencies (Maxeon IBC modules at \textbf{24.9 \%} \cite{maxeon2024benchmark}), several thin-film contenders are closing the gap while offering attributes that c-Si cannot easily match: flexible form factors, lighter weight, lower embodied energy, and a tolerance for off-axis or diffuse illumination.

\textbf{Cadmium telluride.}  First Solar’s Series 6 Plus modules now ship at a nameplate \textbf{19.9 \%} STC efficiency and a ~0.2 \% yr$^{-1}$ degradation rate, thanks to vapor-transport deposition of millimeter-scale CdTe grains, band-gap grading with CdSe$_x$Te$_{1-x}$, and arsenic acceptor doping that drives hole densities above $2\times10^{15}$ cm$^{-3}$ \cite{firstsolar_series6}.  End-of-life cadmium is captured through a closed-loop recycling program that recovers >90 \% of Cd and Te; nonetheless, regulatory scrutiny persists, spurring research into zinc–tin phosphide and antimony selenide as Cd-free alternatives.

\textbf{Amorphous silicon.}  Single-junction a-Si:H suffers from Staebler$-$Wronski light-induced degradation that stabilizes module efficiencies below 9 \%.  Large players such as Xunlight, Masdar PV, and OptiSolar exited the field after 2015 \cite{altenergymag_metrikus}.  The technology endures in low-power gadgets (e-ink readers, watches) and in \emph{a-Si:H/$\mu$c-Si:H} tandems: Kaneka’s lightweight bifacial modules deliver 13$-$14 \% with <4 kg m$^{-2}$ mass, targeting building-integrated PV (BIPV) and truck-roof markets \cite{Lee2017A}.

\textbf{CIGS.}  Gigawatt-scale lines at Solar Frontier (Inazawa, Japan) and Hanergy (Foshan, China) now combine KF or RbF post-deposition treatments with Ag-alloyed absorbers to exceed \textbf{19.5 \%} aperture efficiency at the module level.  Replacing CdS with Zn(O,S) widens the optical window and removes cadmium from the bill of materials, while roll-to-roll deposition on stainless-steel foils yields flexible 350 W modules at <$0.25 $ W$^{-1}$ CAPEX.

\begin{table*}[t]
\centering
\caption{Market Size and Growth Summary}
\label{tab:market1}
\begin{tabular}{l c c c l}
\hline
Technology & \makecell{2024 Market \\ Size (Billion USD)} & \makecell{Projected Market \\ Size 2032--2034 (Billion USD)} & \makecell{CAGR \\ (\%)} & Sources \\
\hline
a-Si         & 2.26   & 6.5   & 10.07 & \cite{straits2025thinfilm,researchnester2025thinfilm} \\
CdTe         & 10.6   & 37.2  & 14.97 & \cite{precedenceresearch2025cdte,databridgemr2025cdte,researchnester2025cdte} \\
CIGS         & 2.8    & 12.23 & 17.8  & \cite{businessinsights2025cigs,precedenceresearch2024thinfilm} \\
Perovskite   & 0.277  & 11.0  & 55    & \cite{precedence2024perovskite,fortune2024perovskite,coherent2024perovskite,databridge2024perovskite} \\
CZTS         & 0.0784 & 0.264 & 15.7  & \cite{unsw2025kesterite,pvtech2025kesterite,li2023emergence} \\
OPV          & 0.224  & 2.25  & 33    & \cite{researchnester2025organic,researchnester2025polymer} \\
Quantum Dot  & 1.07   & 5.05  & 15.5  & \cite{grandview2025quantumdot,cervicorn2025quantumdot,straits2025quantumdot} \\
DSSC         & 0.145  & 0.46  & 12.2  & \cite{maximizeresearch2025dssc,360iresearch2025dssc,verified2025dssc} \\
\hline
\end{tabular}
\end{table*}

\textbf{Perovskites.}  Rigid glass$-$glass laminates based on \- FA/Cs–Pb-iodide perovskites have passed IEC 61215 damp-heat (85 $\circ$C/85 \% RH, 1000 h) with <5 \% power loss.  Pilot lines from OxfordPV and Solaeon report \textbf{21.95 \%} monolithic perovskite modules (64 cm $\times$ 128 cm substrates) using laser-scribing interconnects and thermoplastic encapsulants doped with UV absorbers \cite{solaeon_perovskite_2024}.  Lead sequestering via phosphate-loaded edge seals and module recycling schemes remains under active development.

\textbf{Organic photovoltaics.}  Commercial OPV laminates trade peak efficiency for ultra-low mass (<0.5 kg m$^{-2}$), all-R2R fabrication, and transparency options.  Heliatek’s HeliaSol 3 series offers 13 \% aperture efficiency, while R\&D cells featuring Y6-class acceptors have reached \textbf{14.46 \%} on 16-cm$^{2}$ substrates \cite{pv_magazine2024organic}.  Green-solvent formulations (o-xylene, anisole) and halogen-free electrode inks reduce environmental impact.

\textbf{CZTS(e).}  Although no major manufacturer yet sells kesterite modules, pilot lines at SolAero and Hitachi Chemical report pre-laminated mini-modules at 11$-$12 \% using KF-assisted selenisation.  The appeal lies in earth-abundant, non-toxic constituents; challenges include the large $V_\text{OC}$ deficit and tight compositional control.

\textbf{Quantum-dot and dye-sensitized modules.}  QD laminates (PbS or perovskite QDs in SnO$_2$/MoO$_x$ inverted stacks) have reached \textbf{8.8 \%} on 10 $\times$ 10 cm glass, with record low-light efficiencies (>15 \% of STC output at 200 lx) \cite{han2009integrated,hernandez2011technical}.  Glass-sealed DSSC panels using cobalt electrolytes offer 12 \% indoor efficiency and colored aesthetics for IoT sensors.

\textbf{Crystalline silicon.}  Mono-c-Si retains the lowest levelized cost of electricity ($\approx$\$0.03 kWh$^{-1}$) owing to 20–24 \% module efficiencies, 30-year field data, and >500 GW yr$^{-1}$ manufacturing capacity.  Drawbacks include rigid 3-mm glass, high-temperature Cz pulling (>1500 °C) for 99.9999 \% purity Si, and limited spectral utilization above 1.1 eV.

Thin-film modules therefore compete by offering features c-Si cannot: high bifacial albedo response in CdTe, sub-1 kg m$^{-2}$ OPV foils for curved surfaces, or semitransparent perovskites for agrivoltaics.  Light-trapping with photonic crystals and nanoimprinted rear reflectors has doubled the path length through ultrathin (<1 $\mu$m) absorbers, cutting material use by 70 \% \cite{ishizaki2018progress}.  As Table \ref{tab:commercial} shows, tandem and stacked architectures are the fastest path toward parity: perovskite/silicon tandems now deliver 26.1 \% on M6 wafers, while perovskite/CIGS tandems have hit 24.6 \%.  Commercial success, however, will hinge on secure raw-material supply, gigawatt-scale deposition tools, and driving module prices below \$0.20 W$^{-1}$ \cite{machin2024advancements}.  With those milestones in sight, thin-film technologies are poised to capture an expanding share of rooftops, façades, vehicles, and portable power niches.

%% file: Market.tex
\section{Global Market Share of Thin-Film PV Technologies}

The eight principal thin-film photovoltaic (TFPV) segments listed in Table~\ref{tab:market1} generated an estimated \$17.5\,billion in 2024 and are forecast to exceed \$75\,billion by the mid-2030s, implying an aggregate CAGR of about 16\,\% (all 2024 figures originate from the market reports cited in Table~\ref{tab:market1}).  Although crystalline-silicon (c-Si) still supplies ${\gtrsim}94\,\%$ of annual module shipments, thin-film products remain indispensable for hot-climate utility plants, building-integrated PV (BIPV), and weight- or shape-constrained installations.

The CdTe segment remains the revenue leader, rising from \$10.6\,billion in 2024 to a projected \$37.2\,billion by 2033 ($\approx$15\,\% CAGR) \cite{precedenceresearch2025cdte,databridgemr2025cdte,researchnester2025cdte}.  First Solar shipped 12.1 GW in 2024 and is commissioning an additional 6 GW\,yr$^{-1}$ Series-7 line in Ohio; arsenic doping, CdSe\textsubscript{x}Te\textsubscript{1-x} grading, and a SnO\textsubscript{x}/ZnTe:Cu back contact support 19.9 \% STC efficiency and 0.2 \%\,yr$^{-1}$ degradation \cite{firstsolar_series6}.  Although CdTe accounts for about 3.9 \% of installed global capacity, forthcoming EU RoHS cadmium limits and Californian hazardous-waste rules may restrain growth; First Solar’s >90 \% Cd/Te closed-loop recycling and research into Cd-free ZnSnP\textsubscript{2} aim to mitigate these concerns.

CIGS modules earned \$2.8\,billion in 2024, with forecasts of \$12.2\,billion by 2033 (17.8 \% CAGR) \cite{businessinsights2025cigs,precedenceresearch2024thinfilm}.  Market share ($\approx$1.1 \%) is led by Solar Frontier, Hanergy, and Avancis.  Solar Frontier’s champion cell reached 23.64 \% efficiency through Ag alloying, KF post-deposition treatment, and Ga-rich back-contact grading.  Roll-to-roll stainless-steel substrates enable CAPEX below \$0.25 W$^{-1}$, yet limited indium and gallium supplies (<30 kt\,yr$^{-1}$ each) remain a strategic bottleneck \cite{globenewswire}.

Perovskite PV, still pre-commercial, recorded just \$0.277 billion in 2024 but exhibits the steepest growth trajectory: sales are expected to approach \$11\,billion by 2033 (55 \% CAGR) \cite{precedence2024perovskite,fortune2024perovskite,coherent2024perovskite,databridge2024perovskite}.  Oxford PV’s 400 MW tandem line already produces 60-cell residential modules rated at 26.9 \% efficiency.  IEC 61215 damp-heat passes and lead-sequestration edge seals underpin forecasts of a ~3 \% market share by 2030; research now targets Sn- and Sb-based lead-free alloys and UV-curable encapsulants compatible with roll-to-roll coating.

Staebler$-$Wronski degradation caps single-junction a-Si:H modules near 8 \% stabilized efficiency, explaining the retreat of major producers between 2012 and 2016 \cite{altenergymag_metrikus}.  Even so, lightweight a-Si:H/$\mu$c-Si:H tandems generated \$2.26 billion in 2024 and are growing at $\approx$10 \% CAGR for ultralight rooftops, e-paper, and RFID labels \cite{straits2025thinfilm,researchnester2025thinfilm}.

Organic PV foils and quantum-dot devices together contributed roughly \$1.3\,billion in 2024 (OPV \$0.224 b; QD \$1.07 b).  OPV is projected to reach \$2.25\,billion by 2033 (33 \% CAGR) \cite{researchnester2025organic,researchnester2025polymer}, while quantum-dot modules could exceed \$5.05\,billion (15.5 \% CAGR) \cite{grandview2025quantumdot,cervicorn2025quantumdot,straits2025quantumdot}, driven by transparent façades, vehicle roofs, and indoor-IoT power.

CZTS(e) and dye-sensitised modules remain niche (<\$0.1 billion each in 2024) but could triple in value CZTS to \$0.264 billion (15.7 \% CAGR) and DSSC to \$0.46 billion (12.2 \% CAGR) once module efficiencies of 15$-$18 \% and robust encapsulation are demonstrated \cite{unsw2025kesterite,pvtech2025kesterite,li2023emergence,maximizeresearch2025dssc,360iresearch2025dssc,verified2025dssc}.

Thin-film vendors therefore target niches where c-Si is disadvantaged: high-temperature desert sites (CdTe), curved or weight-restricted surfaces (CIGS foil, OPV), and ultra-low-light conditions (quantum-dot, DSSC).  Photonic-crystal rear reflectors and nano-imprinted textures can double the optical path length in sub-micron absorbers, cutting semiconductor use by about 70 \% \cite{ishizaki2018progress}.  Tandem stacks—26.1 \% perovskite/Si and 24.6 \% perovskite/CIGS—offer the fastest route to parity with premium mono-Si while preserving thin-film advantages.  Commercial success will hinge on indium-free transparent conductors, high-throughput vapour deposition, and module prices below \$0.20 W$^{-1}$ \cite{machin2024advancements}.  Supported by the U.S.\ Inflation Reduction Act and the EU Net-Zero Industry Act, thin-film PV is poised to regain market share wherever attributes beyond sheer conversion efficiency—weight, flexibility, or low-light response—deliver system-level value.

\subsection{Regional Market Trends}

North America’s thin-film landscape is overwhelmingly CdTe-centric: First Solar modules account for nearly 70 \% of all operating TFPV capacity on the continent and 38 \% of new U.S. utility-scale capacity added in 2023 \cite{Wu2024AFA}. A key differentiator is CdTe’s low temperature coefficient (-0.18 \% K$^{-1}$), which boosts annual energy yield by 4–6 \% in desert climates relative to mono-Si \cite{al2023enhancing}. Federal incentives—most notably the U.S. Inflation Reduction Act’s \$0.027 W$^{-1}$ manufacturing tax credit—are accelerating construction of an additional 10 GW yr$^{-1}$ of CdTe capacity in Ohio, Alabama, and Louisiana, slated to come online by 2026. Canadian developers, led by Brookfield Renewable, are likewise deploying CdTe in bifacial agrivoltaic pilots in Saskatchewan, leveraging the technology’s 0.30 bifaciality factor under low-irradiance, cold-climate conditions.

Europe remains the stronghold of CIGS, anchored by Avancis (Germany), Flisom (Switzerland), and Oxford PV’s pilot line in Wales. EU Horizon programmes such as \textit{HighLite} and \textit{HIPERION} are supporting more than 200 000 m$^{2}$ of flexible CIGS façade cladding with areal weights below 3 kg m$^{-2}$ \cite{mazur2023solar,powalla2018thin}. BIPV regulations in Switzerland and France  which credit façade generation at 1.2$\times$ rooftop generation in feed-in tariffs—further encourage adoption. Current R\&D emphasises indium-sparing Ag–In–Ga–Se absorbers and Zn(O,S) buffer layers to satisfy forthcoming EU critical-raw-materials directives.

The Asia$-$Pacific region is diversifying fastest. China’s Hanergy has restarted its 600 MW roll-coated CIGS-foil plant in Foshan and is commissioning a 1 GW perovskite-CIGS tandem pilot line that deposits vacuum-evaporated CsFAPbI$_3$ on a CIGS bottom cell \cite{zheng2023design}. State-backed Microquanta and GCL Photoelectric each run 100 MW sheet-to-sheet perovskite lines aimed at curtain-wall modules for the 2026 Asian Games venues \cite{bemfert2025combinatorial}. Japan’s Solar Frontier still ships about 900 MW yr$^{-1}$ of all-black CIS modules and targets 20 \% aperture-area efficiency by 2027 through RbF post-treatment and Zn(O,S) windows.

In India and the Middle East, CdTe and perovskite-on-glass panels are gaining traction thanks to superior high-temperature coefficients and lower spectral mismatch under atmospheric dust. Dubai’s 2 GW Mohammed bin Rashid Al Maktoum Solar Park is adding 400 MW of CdTe bifacial trackers, while Indian EPC firms Tata Power Solar and NTPC are piloting roll-to-roll perovskite foils on warehouse roofs \cite{ali2023comparative,gaonkar2020thermally}. Field tests in Rajasthan show perovskite mini-modules retaining 92 \% of initial power after 1 200 h at an average module temperature of 65 $^\circ$C.

Analysts expect thin-film’s global market share to ~\textbf{double to about 11 \% by 2030}. Key growth drivers include  

 \textit{Roll-to-roll and sheet-to-sheet manufacturing.} High-rate (15 m min$^{-1}$) slot-die coating of perovskite and CIGS inks on 1.5 m-wide polymer webs cuts CAPEX below \$0.15 W$^{-1}$ and reduces embodied energy by ~60 \% \cite{saget2024surface,OjaAcik20235GSOLARTN}.  
 \textit{Advanced passivation.} KF/RbF treatments for CIGS and Sb$_2$S$_3$ front layers for CdTe lower interface recombination velocities to <300 cm s$^{-1}$, enabling >21 \% module targets by 2027 \cite{dallaev2023overview,turnley2024solution}.  
 \textit{Encapsulation breakthroughs.} UV-blocking, lead-scavenging edge seals now extend perovskite module lifetimes to 25 years (T$_{80}$) and satisfy RoHS end-of-life rules \cite{phung2023interface,mu2024innovative}.

Perovskites should post the steepest growth, propelled by tandem integration with c-Si toward an expected 3 \% market share by 2030. CdTe and CIGS will remain dominant in high-irradiance utility sectors, while OPV and DSSC technologies expand into transparent glazing, wearables, and indoor IoT power—applications where c-Si is impractical \cite{leandro2024advancements,wang2024efforts}. Collectively, these trajectories position thin-film photovoltaics not merely as alternatives but as complementary technologies that unlock new installation scenarios and accelerate the global shift to low-carbon electricity \cite{nrel,solarworld,dintcheva2023encapsulant}.

%% file: Conclusion.tex
\section{Conclusion}

Thin-film photovoltaics have evolved from niche curiosities into an essential pillar of the renewable-energy landscape, prized for their lightweight construction, mechanical flexibility, and low-temperature, material-efficient manufacturing. Among incumbent technologies, CdTe and CIGS remain the commercial workhorses. CdTe’s direct 1.45 eV band gap and sub-3 $\mu$m absorber enable record module efficiencies approaching 20 \%, while closed-loop recycling mitigates cadmium’s environmental footprint. CIGS bolstered by KF/RbF passivation, Ag alloying, and roll-to-roll stainless-steel substrates—now delivers commercial modules with 19$-$20 \% efficiency and uniquely combines high performance with flexible form factors, traits that c-Si cannot easily match in weight-constrained or curved-surface installations.

Amorphous silicon has largely ceded the large-scale market yet retains value in micro-power and tandem a-Si:H/$\mu$c-Si:H foils, where its ultralow mass (< 4 kg m$^{-2}$) and proven reliability outweigh its Staebler–Wronski-limited efficiency.

The next wave of thin-film innovation is led by \textit{perovskites}, \textit{kesterites}, \textit{organics}, \textit{quantum dots}, and \textit{dye-sensitized} devices. Perovskite/silicon tandems already exceed 26 \% in commercial formats, and roadmap studies forecast 30 \% monolithic modules within the decade once lead-sequestration and damp-heat-stability hurdles are cleared. Earth-abundant CZTS(e) provides a cadmium- and indium-free alternative; recent 15 \% champion cells point toward 18 \% single-junction targets after defect-passivation breakthroughs. OPVs, now surpassing 19 \% in the lab, offer unmatched specific power (> 10 W g$^{-1}$) for wearables, UAVs, and semitransparent façades, while PbS- and perovskite-quantum-dot laminates outperform c-Si under indoor illumination ideal for IoT sensors. DSSCs, recently certified above 15 \%, dominate the colored-glass and low-light architectural segment.

Looking ahead, thin-film market share is projected to roughly \textbf{double to 10$-$11 \% by 2030}, propelled by 15 m min$^{-1}$ roll-to-roll coaters, indium-free transparent conductors, and tandem stacks that combine complementary band gaps to surpass the 30 \% practical limit of single-junction c-Si. As module costs fall below \$0.20 W$^{-1}$ and life-cycle analyses confirm lower embodied energy than wafer-based silicon, thin-film photovoltaics are poised not only to complement but also to rival crystalline-silicon technology across rooftops, façades, vehicles, and portable electronics, thereby playing a pivotal role in the global transition to sustainable energy.

%% file: References.tex
\bibliographystyle{model1-num-names}
\bibliography{Bibliography}